\documentclass{JHEP3} % 10pt is ignored!

\usepackage{graphicx}
\usepackage{amsmath, amsfonts}
\usepackage{epsfig}
\newcommand{\non}{\nonumber\\}

\newcommand{\be}{\begin{equation}}
\newcommand{\ee}{\end{equation}}
\newcommand{\bea}{\begin{eqnarray}}
\newcommand{\eea}{\end{eqnarray}}
\newcommand{\ba}[1]{\begin{array}{#1}}
\newcommand{\ea}{\end{array}}

\newcommand{\bm}[1]{\mbox{\boldmath${#1}$}}

\title{Meson supercurrents and the Meissner effect in the Sakai-Sugimoto model}

\author{Anton Rebhan, Andreas Schmitt, Stefan A. Stricker\\
Institut f\"{u}r Theoretische Physik, Technische Universit\"{a}t Wien, 1040 Vienna, Austria\\ 
\\
 \email{rebhana@hep.itp.tuwien.ac.at}\\
 \email{aschmitt@hep.itp.tuwien.ac.at}\\
  \email{stricker@hep.itp.tuwien.ac.at}}

\abstract{The Sakai-Sugimoto model provides a holographic description for chiral symmetry breaking. We use this model
to investigate chirally broken phases in an external magnetic field at finite 
isospin and baryon chemical potentials. The equations of motion for the bulk gauge fields
are solved analytically and the free energy is computed from the Yang-Mills and Chern-Simons contributions to the D8 brane action.
In the case of a neutral pion condensate, a magnetic field is found to induce 
nonzero gradients of the Goldstone boson fields corresponding to meson supercurrents. 
A charged pion condensate, on the other hand, expels the magnetic field due to the Meissner effect. 
Upon comparing the Gibbs free energies of these two phases we find that the rotation of the
chiral condensate into a charged pion condensate for finite isospin chemical potentials is partially undone by switching on 
a magnetic field, and we determine the critical magnetic field which
removes the charged pion condensate in a first-order phase transition.}


\begin{document}

%%%%%%%%%%%%%%%%%%%%%%%%%%%%%%%%%%%%%%%%%%%%%%%%%%%%%%%%%%%%%%%%%%%%%%%%%%%%%%%%%%%%%%%
\section{Introduction}
\label{intro}
%%%%%%%%%%%%%%%%%%%%%%%%%%%%%%%%%%%%%%%%%%%%%%%%%%%%%%%%%%%%%%%%%%%%%%%%%%%%%%%%%%%%%%%

Matter in large magnetic fields occurs in nature in the interior of neutron stars \cite{Duncan:1992hi}, or can be created in the laboratory 
from noncentral collisions of heavy ions \cite{Kharzeev:2007jp}.  In both contexts it is important to understand the 
response of strongly-interacting quark matter, ultimately described by Quantum Chromodynamics (QCD), 
to an external magnetic field. And in both contexts the interplay of the magnetic field with chiral properties of the matter, in particular with 
chiral symmetry breaking, are crucial. While heavy-ion collisions probe the regime of 
the chiral phase transition at large temperatures and small chemical potentials, highly magnetized neutron stars (``magnetars'') 
are composed of dense and comparably cold matter, which also may be in a phase with broken chiral symmetry. In dense quark matter, 
chiral symmetry can be broken by the usual chiral condensates $\langle\bar{\psi}\psi\rangle$ or, in a three-flavor system at 
sufficiently large densities, by diquark
condensates $\langle\psi\psi\rangle$ in the color-flavor locked state \cite{Alford:1998mk}. In this paper we shall consider a two-flavor system 
at finite baryon and isospin chemical potential in the strong-coupling regime at large $N_c$, which may form different kinds of chiral condensates 
$\langle\bar{\psi}\psi\rangle$ depending on the values of temperature, the chemical potentials, and the external magnetic field.

It has been shown using a Nambu-Jona-Lasinio (NJL) model that a magnetic field can act as a catalyst for chiral symmetry breaking 
\cite{Gusynin:1994re,Gusynin:1995nb}, see also \cite{Klevansky:1989vi,Klevansky:1992qe}. 
Also chiral perturbation theory has been used to study the effect 
of magnetic fields  \cite{Agasian:2001ym,Cohen:2007bt}, recently for instance in the context of the deconfinement and chiral phase transitions
in Refs.\ \cite{Agasian:2008tb} and \cite{Fraga:2008qn,Mizher:2008hf}, respectively. All these studies are restricted to the vacuum, i.e., they 
are done for the case of vanishing chemical potentials. Dense matter with nonvanishing chemical potentials
in a magnetic field has been studied in the context of color superconductivity \cite{Ferrer:2005vd,Fukushima:2007fc,Noronha:2007wg}, which, 
due to Goldstone boson currents and the axial anomaly, can be ferromagnetic \cite{Son:2007ny}. In this paper, we use the holographic
model by Sakai and Sugimoto \cite{Sakai:2004cn,Sakai:2005yt} at nonzero isospin and baryon chemical potentials
to study the effect of a magnetic field on chirally 
broken phases.\footnote{For effects of magnetic fields in other holographic
models of strongly coupled gauge theories with flavor degrees of freedom see e.g.\ Refs.~\cite{Filev:2007gb,Albash:2007bk,Erdmenger:2007bn}.}

Holographic models have been used extensively to study the strong-coupling limit of gauge theories since the conjecture of the AdS/CFT 
correspondence \cite{Maldacena:1997re,Gubser:1998bc,Witten:1998qj}, i.e., the duality between (the supergravity approximation of) 
string theories and 
(the strong-coupling limit of) conformal supersymmetric gauge theories; for a review see Ref.\ \cite{Aharony:1999ti}. The original and 
most prominent theory under investigation has been ${\cal N}=4$ super Yang-Mills theory which lives on the 3+1 dimensional boundary of 
AdS$_5$$\times$S$^5$, and which is dual to type-IIB string theory living in this ten-dimensional space. While this supersymmetric 
gauge theory shares several properties with QCD, the differences to QCD are significant and thus it mainly serves as a model whose results 
should be compared with, not interpreted as, results from QCD. Therefore, a lot of effort has been spent to develop a gravity dual 
of QCD. Such a dual is unknown at present. Besides a ``bottom-up'' approach to AdS/QCD \cite{Son:2003et,Erlich:2005qh}, a promising
model for capturing essential features of QCD is the Sakai-Sugimoto model, developed to study the physics of chiral symmetry breaking and 
meson spectra \cite{Sakai:2004cn,Sakai:2005yt}. In contrast to the original AdS/CFT correspondence, but building upon an early proposal by
Witten \cite{Witten:1998zw}, the Sakai-Sugimoto model deals with type-IIA 
string theory and a five-dimensional dual gauge theory, where by means of a compactified extra dimension  
supersymmetry is completely broken. This extra dimension separates massless fermions of right- and left-handed chirality, which are located
on the intersections of D8 and $\overline{\rm D8}$ branes
with $N_c$ D4 branes. As all other known gravity duals, the model is dual to the large-$N_c$ limit of the field theory, and the 
simple supergravity approximation, used in this paper as well as in most previous studies, corresponds to large values 
of the 't Hooft coupling.

With nonzero chemical potentials and a magnetic field,
we shall find meson supercurrents and the Meissner effect in the chirally broken phases. Both phenomena are best understood 
as an analogy to (weak-coupling) superfluidity or superconductivity. For instance, 
a charged pion condensate
of the form $\langle\bar{d}\gamma_5 u\rangle$ can be viewed as Cooper pairing of two different fermion species,
here an anti-down-quark and an up-quark. 
In general, Cooper pairing of two fermion species with chemical potentials $\mu_1$ and $\mu_2$ 
takes place at a common Fermi surface given by $\bar{\mu}=(\mu_1+\mu_2)/2$. A mismatch in chemical potentials $\delta\mu=(\mu_1- \mu_2)/2$ 
induces a ``stress'' on the pairing in trying to move the two Fermi surfaces apart. 
For not too large values of $\delta\mu$, the system can sustain the stress and the 
densities of the two fermion species are (at zero temperature) ``locked'' together, 
i.e., the difference in densities $\delta n=n_1-n_2$ vanishes. 
For larger values of $\delta\mu$, and before completely breaking the condensate, the system may respond to the stress 
by leaving some, but not all, fermions around the Fermi surfaces unpaired, allowing for a nonzero $\delta n$. The resulting 
state breaks rotational invariance, and it may even break translational invariance by giving rise to a crystalline structure. 
Anisotropic pion condensates in nuclear matter have been discussed a long time ago 
\cite{1973JETP...36.1052M,Sawyer:1972cq,Scalapino:1972fu,Baym:1973zk};
crystalline structures of the superfluid order 
parameter are well-known in condensed matter physics \cite{LarkinOvchinnikov,FuldeFerrell} as well as in dense quark matter 
\cite{Alford:2000ze}, and also have been 
discussed in the context of chiral condensates \cite{Schnetz:2004vr}, see also \cite{He:2006tn}. In either case, be it 
in a homogeneous manner or in a complicated crystalline structure, this unconventional pairing induces nonzero  
``supercurrents'' in the system, see for instance Refs.\ \cite{Kryjevski:2008zz,Schafer:2005ym,Schmitt:2008gn}.  
These supercurrents are cancelled by counter-propagating 
currents, typically coming from unpaired fermions, such that the net current in the system vanishes.
  
In the case of a pion condensate of the form $\langle\bar{d}\gamma_5 u\rangle$, $\bar{\mu}$ and $\delta\mu$ correspond to
the isospin, $\mu_I$, and baryon, $\mu_B$, chemical potential, respectively. Consequently, one might expect anisotropic
pairing upon increasing the ``mismatch'' $\mu_B$. And, corresponding to the above $\delta n$, a nonzero baryon number $n_B$
is expected. In the Sakai-Sugimoto model, a finite baryon number is taken into account via the Chern-Simons term. 
Localized baryons can be described by instantons of the
effective gauge theory of the flavor branes \cite{Nawa:2006gv,Hata:2007mb,Hashimoto:2008zw,Nawa:2008uv,Seki:2008mu} corresponding to 
chiral skyrmions, which in the ground state form
crystals rather than a liquid \cite{Rozali:2007rx,Kim:2007zm,Kim:2007vd}.
However, we are interested in a homogeneous 
distribution of baryon (and isospin) density. 
It turns out that this can be achieved by
a nonzero magnetic field in the model \cite{Bergman:2008qv,Thompson:2008qw},
which is anyway of interest in the context of neutron star
physics. 

A magnetic field, however, is expelled from the charged pion condensate because a condensate of charged bosons (be it Cooper pairs or, in our 
case, Goldstone bosons) acts as a superconductor and thus exhibits a Meissner effect.\footnote{Holographic models of superconductors and superfluids
have recently been investigated in Refs. \cite{Hartnoll:2008vx,Gubser:2008wv,Basu:2008st,Herzog:2008he,Ammon:2008fc,Basu:2008bh}, 
see Ref.\ \cite{Hartnoll:2008kx} for a discussion of the Meissner effect.} Accordingly, we shall find the above expectations
of a supercurrent and nonzero baryon number not realized in the charged pion condensate which remains unmodified for (not too large) 
magnetic fields. A meson supercurrent as well as nonzero baryon (and isospin) numbers occur, for nonzero magnetic field, 
instead in the phase with a neutral pion condensate.
This phase is energetically preferred over the charged pion condensate
beyond a critical magnetic field which we shall compute.

Our paper is organized as follows. In Sec.\ \ref{sec:model} we briefly recapitulate the Sakai-Sugimoto
model. We discuss our ansatz for solutions in the presence
of baryon and isospin chemical potentials and a magnetic field
and derive the equations of 
motion and the free energy for the chirally broken phases in Sec.\ \ref{confinedphase}. The main part of the paper is Sec.\ \ref{sec:broken}.
In this part we first discuss how to incorporate different chiral condensates into the model, see Sec.\ \ref{sec:rotate}. In Sec.\ \ref{sec:solve}
we solve the equations of motion for the sigma and the charged pion phase and compute their free energies. In particular, we discuss the 
Meissner effect in Sec.\ \ref{sec:meissner}. The results are used to discuss 
the currents and number densities in these phases in Sec.\ \ref{sec:properties1}. Finally, we compare their free energies
to discuss the resulting phase diagram in Sec.\ \ref{sec:compare}, and we give our conclusions
in Sec.\ \ref{sec:summary}.

%%%%%%%%%%%%%%%%%%%%%%%%%%%%%%%%%%%%%%%%%%%%%%%%%%%%%%%%%%%%%%%%%%%%%%%%%%%%%%%%%%%%%%%%%%%%
\section{The model}
\label{sec:model}
%%%%%%%%%%%%%%%%%%%%%%%%%%%%%%%%%%%%%%%%%%%%%%%%%%%%%%%%%%%%%%%%%%%%%%%%%%%%%%%%%%%%%%%%%%%%%

In this section, the basic equations of the Sakai-Sugimoto model shall be summarized. They will be needed in the subsequent sections.
For more details about the setup of the model see for instance the original papers by Sakai and Sugimoto \cite{Sakai:2004cn,Sakai:2005yt}. 

%%%%%%%%%%%%%%%%%%%%%%%%%%%%%%%%%%%%%%%%%%%%%%%%%%%%%%%%%%%%%%%%%%%%%%%%%%%%%%%%%%%%%%%%%%%%
\subsection{Geometry of confined and deconfined phases}
%%%%%%%%%%%%%%%%%%%%%%%%%%%%%%%%%%%%%%%%%%%%%%%%%%%%%%%%%%%%%%%%%%%%%%%%%%%%%%%%%%%%%%%%%%%%

The bulk background geometry is given by the ten-dimensional supergravity
description of $N_c$ D4 branes in type-IIA superstring theory compactified
on a circle. There are two different solutions for the metric, realized in two different temperature regimes. 
The transition from one to the other is interpreted as the deconfinement phase transition. 
Similar to the original AdS/CFT setting at finite temperature \cite{Witten:1998zw}, the deconfined phase has a black hole which is absent 
in the confined phase. 

The (euclidean) metric of the confined phase is given by 
\cite{Kruczenski:2003uq}
\be \label{dsconf}
ds^2_{\rm conf}= \left(\frac{u}{R}\right)^{3/2}[dt^2+\delta_{ij}dx^idx^j+f(u)dx_4^2]+
\left(\frac{R}{u}\right)^{3/2}\left[\frac{du^2}{f(u)}+u^2d\Omega_4^2\right] \, .
\ee
Here, $d\Omega_4^2$ is the metric of a four-sphere, and $R$ is the curvature 
radius of the background which is related to the string coupling 
$g_s$ and the string length $\ell_s$ via
\be \label{R3}
R^3=\pi g_s N_c \ell_s^3 \, .
\ee
A crucial feature of the model is
the compactified dimension $x_4$ which has a radius which we can parametrize through the Kaluza-Klein mass 
\be \label{MKK}
M_{\rm KK}=\frac{3}{2}\frac{u_{\rm KK}^{1/2}}{R^{3/2}} \, , \qquad x_4\equiv x_4 +2\pi/M_{\rm KK} \, .
\ee
This breaks supersymmetry completely by giving Kaluza-Klein masses to
the adjoint fermions of the dual gauge theory and the analogue of
thermal masses to the adjoint scalars, leaving only gauge bosons in
the spectrum of the
low-energy limit as the latter are protected by gauge symmetry
\cite{Witten:1998zw}.
The point $u=u_{\rm KK}$ is the tip of the cigar-shaped subspace spanned by $x_4$ and the 
holographic coordinate $u$, and  
\be
f(u)\equiv 1-\frac{u_{\rm KK}^3}{u^3} \, .
\ee
%The string coupling can be expressed in terms of the Kaluza-Klein mass as 
%\be \label{coupling}
%g_s = \frac{\lambda}{2\pi M_{\rm KK}\ell_s N_c} \,  ,
%\ee
%where $\lambda$ is the dimensionless 't Hooft coupling of the four dimensional field theory.   
The subspace spanned by the euclidean time $t$ and the coordinate $u$ is cylinder-shaped, with the circumference given by the inverse
temperature, $t\equiv t + 1/T$. In the deconfined phase the coordinates $t$ and $x_4$ interchange their roles, i.e., now the subspace 
spanned by $x_4$ and $u$ is cylinder-shaped while the subspace spanned by $t$ and $u$ is cigar-shaped. In this case, the metric is
\be \label{dsdeconf}
ds^2_{\rm deconf}= \left(\frac{u}{R}\right)^{3/2}[\tilde{f}(u)dt^2+\delta_{ij}dx^idx^j+dx_4^2]+
\left(\frac{R}{u}\right)^{3/2}\left[\frac{du^2}{\tilde{f}(u)}+u^2d\Omega_4^2\right] \, ,
\ee
where temperature is related to the tip of the cigar-shaped $t$-$u$ space $u_T$ via
\be \label{T}
T=\frac{3}{4\pi}\frac{u_T^{1/2}}{R^{3/2}} \, , 
\ee
and 
\be
\tilde{f}(u)\equiv 1-\frac{u_T^3}{u^3} \, .
\ee
The deconfinement phase transition is located at a critical temperature $T=T_c$ where the free energies corresponding to the two phases 
are identical. This occurs at $u_{\rm KK}=u_{T_c}$ and thus $T_c=M_{\rm KK}/(2\pi)$. This critical temperature is independent 
of the chemical potential. Consequently, the model predicts a horizontal phase transition line in the $T$-$\mu_B$ plane,
in accordance with expectations from QCD at infinite number of colors $N_c$ \cite{McLerran:2007qj}.

The supergravity prescription depends on having
the background weakly curved compared to the string scale.
This is the case for large four-dimensional
't Hooft coupling \cite{Kruczenski:2003uq}
\be\label{coupling}
\lambda=g_{\rm YM}^2 N_c=\frac{g_5^2N_c}{2\pi M_{\rm KK}^{-1}}\gg1 \, ,
\ee
where the five-dimensional gauge coupling $g_5$ is given by
$g_5^2=(2\pi)^2g_s \ell_s$.

The Kaluza-Klein mass sets the energy scale below which the dual field theory is effectively four-dimensional. 
For large 't Hooft coupling,
this scale is of the same order as the mass gap of the field theory; only
for small $\lambda$, where string corrections become important, does one
have duality with non-supersymmetric large-$N_c$ QCD in four dimensions. 
However, there is already ample evidence 
that the limit of large 't Hooft coupling,
where supergravity calculations are meaningful, does provide
a useful tool for unravelling certain nonperturbative features
of QCD.

Sakai and Sugimoto \cite{Sakai:2004cn} added $N_f$ pairs of D8 and $\overline{\rm D8}$ branes which are transverse to the circle along $x_4$.
The intersections of these branes with the D4 branes carry
massless fermions in the fundamental representation of the color group
of opposite chirality, which are interpreted as massless quarks
of QCD.
%, where $N_f$ shall be interpreted 
%as the number of flavors, as will become clear below. 
As long as $N_f\ll N_c$, the D8/$\overline{\rm D8}$  branes can be treated as probe
branes, i.e., the backreaction on the background geometry is neglected.
%We shall 
%treat these branes as probe branes, i.e., we shall always neglect the backreaction to the background geometry. This approximation 
%requires $N_f\ll N_c$. 
Below we shall take $N_f=2$. The D8 and $\overline{\rm D8}$ branes extend in all dimensions except for the 
coordinate $x_4$ (whereas the D4 branes extend in the $t$, $x_i$, $i=1,\ldots ,4$ directions). The induced metrics
on the probe branes in the confined and deconfined backgrounds are 
\begin{subequations}
\bea \label{dsD8conf}
ds_{\rm D8,conf}^2&=& \left(\frac{u}{R}\right)^{3/2}\left(dt^2+\delta_{ij}dx^idx^j\right)+\left(\frac{R}{u}\right)^{3/2}\left[
\frac{v^2(u)}{f(u)} du^2
+u^2d\Omega_4\right] \, , \\
\label{dsD8deconf}
ds_{\rm D8,deconf}^2&=& \left(\frac{u}{R}\right)^{3/2}\left[\tilde{f}(u)dt^2+\delta_{ij}dx^idx^j\right]+\left(\frac{R}{u}\right)^{3/2}\left[
\frac{\tilde{v}^2(u)}{\tilde{f}(u)} du^2 
+u^2d\Omega_4\right] \, ,
\eea
\end{subequations}
where we abbreviated
\be \label{vvtilde}
v(u)\equiv \sqrt{1+f^2(u)\left(\frac{u}{R}\right)^3 (\partial_ux_4)^2} \, , \qquad 
\tilde{v}(u)\equiv \sqrt{1+\left(\frac{u}{R}\right)^3 (\partial_ux_4)^2} \, .
\ee
Here the function $x_4(u)$ gives the embedding of the D8 branes in the $x_4$-$u$ subspace. 

The D4/D8-$\overline{\rm D8}$ setup provides the tools to study not only the deconfinement phase transition but 
also the chiral phase transition. In the $x_4$ direction, the D8 branes are separated from the $\overline{\rm D8}$ branes by a distance $L$. 
The maximal separation 
of the branes is $L=\pi/M_{\rm KK}$ in which case the branes are attached at opposite sides of the circle spanned by $x_4$.
Gauge fields on the D8 and $\overline{\rm D8}$ branes transforming under a local symmetry group $U(N_f)$ induce  
a global symmetry group $U(N_f)$ on the five-dimensional boundary at $u=\infty$. More precisely, 
a gauge symmetry on the D8 branes induces a global symmetry 
at the four-dimensional subspace of the holographic boundary at $x_4=0$, while the gauge symmetry on the $\overline{\rm D8}$ branes induces
a separate global symmetry on the four-dimensional subspace at $x_4=L$. Therefore the total global symmetry can be interpreted
as the chiral group $U(N_f)_L\times U(N_f)_R$. 

So far we have viewed the gauge symmetry on the D8 branes as independent from that on the 
$\overline{\rm D8}$ branes. This is correct if the branes are geometrically separate. For example in the deconfined background, where the 
$x_4$-$u$ subspace is cylinder-shaped, the branes follow straight lines from $u=u_T$ up to $u=\infty$, and thus are disconnected. However, 
it may also be energetically favored for the branes to be connected. In this case, the gauge symmetry reduces to joint rotations,
given by the vectorial subgroup $U(N_f)_{L+R}$. This is exactly the symmetry breaking pattern induced by a chiral condensate
(for a discussion of the chiral condensate itself as an order parameter for chiral symmetry breaking within the Sakai-Sugimoto model
see Refs.\ \cite{Aharony:2008an,Argyres:2008sw}). 
In fact, in the confined phase, where the $x_4$-$u$ subspace is cigar-shaped, the branes {\it must} connect. In other words, 
chiral symmetry is always broken in the confined phase.  Whether the branes are  
disconnected in the deconfined phase depends on the separation scale $L$. For sufficiently large $L$ they are always disconnected, while for 
smaller $L$ the connected phase may be favored for certain temperatures \cite{Aharony:2006da}. In other words, in the former case, 
deconfinement and the chiral 
phase transition are identical while in the latter case they differ and there exists a deconfined but chirally broken phase in the 
$T$-$\mu_B$ plane \cite{Horigome:2006xu}. In this paper, we shall use maximally separated branes, i.e., $L=\pi/M_{\rm KK}$. This simplifies
the treatment since in this case we always have $\partial_u x_4 =0$ because the D8 branes follow geodesics. The case of not maximally separated
planes, more precisely the limit where the radius of the compactified dimension is much larger than the separation 
distance, $1/M_{\rm KK}\gg L$, corresponds to an NJL model on the field theory side \cite{Antonyan:2006vw}. 

Temperature and chemical potentials enter the model in very different ways. As explained above, temperature has a geometric effect 
on the background metric, in particular a black hole forms for sufficiently large $T$. Chemical potentials, however, enter as boundary 
conditions for the gauge fields on the D8 and $\overline{\rm D8}$ branes, i.e., in the subsequent sections we will fix the baryon and isospin 
components of the 
temporal components of the ``right-handed'' and ``left-handed'' gauge fields at the boundary $u=\infty$ by the isospin and baryon 
chemical potentials. 
Analogously, nontrivial boundary values of the spatial components
of the gauge fields have the interpretation of spatial gradients
in chiral condensates, corresponding to supercurrents. 
We shall discuss the gauge field action associated with
the flavor branes in more detail now.

%%%%%%%%%%%%%%%%%%%%%%%%%%%%%%%%%%%%%%%%%%%%%%%%%%%%%%%%%%%%%%%%%%%%%%%%%%%%%%%%%%%%%%%%%%%%
\subsection{Yang-Mills and Chern-Simons action}
%%%%%%%%%%%%%%%%%%%%%%%%%%%%%%%%%%%%%%%%%%%%%%%%%%%%%%%%%%%%%%%%%%%%%%%%%%%%%%%%%%%%%%%%%%%%

The total action for the D8 and $\overline{\rm D8}$
branes is given by the sum of the Dirac-Born-Infeld (DBI) and the Chern-Simons actions. As indicated in the introduction and as 
will become clear below, the Chern-Simons term is necessary to account for nonzero baryon and isospin numbers and, in our ansatz, will be 
proportional to the external magnetic field. For simplicity, we shall expand the DBI action for small gauge fields such that we 
obtain a Yang-Mills contribution instead. This was also done for instance in Ref.\ \cite{Thompson:2008qw}, while other works  
used the full DBI action in a similar context, however for the simpler cases of a one-flavor system without isospin chemical potential 
\cite{Bergman:2008qv} and without currents and magnetic field \cite{Parnachev:2007bc}. Our action takes the form
\be \label{totalS}
S_{\rm D8}= S_{\rm YM} + S_{\rm CS} \, . 
\ee
Here, the Yang-Mills contribution is
\be \label{SYM}
S_{\rm YM} = 2 N_f T_8 V_4\int d^4x\, du \, e^{-\Phi}\sqrt{g}\left(1-\frac{(2\pi\alpha')^2}{4N_f}
 g^{\mu\nu}g^{\rho\sigma} {\rm Tr}[{\cal F}_{\nu\rho}{\cal F}_{\sigma\mu}] \right) \, , 
\ee
where $T_8=1/[(2\pi)^8\ell_s^9]$ is the D8-brane tension, where $\alpha'=\ell_s^2$, and where $V_4=8\pi^2/3$ is the volume of the 
unit 4-sphere. The remaining integrations are done over four-dimensional space-time $t,x_1,x_2,x_3$, and over 
the holographic coordinate $u$. In the confined (deconfined) phase the limits for this integration are $u_{\rm KK} (u_T)<u<\infty$, and 
the factor 2 on the right-hand side of Eq.\ (\ref{SYM}) accounts for integration over D8 and $\overline{\rm D8}$ branes. 
In this section, all expressions are thus valid for both confined and deconfined phase, which differ, besides the integration limits for $u$,
by the metric $g$. 
The dilaton is $e^{\Phi}=g_s(u/R)^{3/4}$, and the trace is taken over the 
internal $U(2)$ space (from now on $N_f=2$). Our convention for the field strength tensor is
\be \label{Fmunu}
{\cal F}_{\mu\nu}= \partial_\mu {\cal A}_\nu -\partial_\nu {\cal A}_\mu -i[{\cal A}_\mu,{\cal A}_\nu] \, , 
\ee
where $\mu,\nu=0,1,2,3,u$, and where ${\cal A}_\mu$ is the $U(2)$ gauge field. It is convenient to separate the 
$U(1)$ part from the gauge fields and field strengths,
\be \label{convention}
{\cal A}_\mu = \frac{\hat{A}_\mu}{2} {\bf 1} + \frac{A^a_\mu}{2}\tau_a \, , \qquad {\cal F}_{\mu\nu} = \frac{\hat{F}_{\mu\nu}}{2} {\bf 1} + 
\frac{F_{\mu\nu}^a}{2}\tau_a \, , 
\ee
where $a=1,2,3$ and $\tau_a$ are the Pauli matrices.  With these conventions we have 
\be
\hat{F}_{\mu\nu}=\partial_\mu \hat{A}_\nu -\partial_\nu \hat{A}_\mu \, , \qquad 
F_{\mu\nu}^a=\partial_\mu A_\nu^a -\partial_\nu A_\mu^a +A_\mu^bA_\nu^c\epsilon_{abc} \, .
\ee 
The Chern-Simons contribution in Eq.\ (\ref{totalS}) is \cite{Hata:2007mb}
\bea \label{SCS}
S_{\rm CS}&=&-i\frac{N_c}{12\pi^2}\int\left\{\frac{3}{2}\hat{A}{\rm Tr}[F^2]+\frac{1}{4}\hat{A}\hat{F}^2+\frac{1}{2}d\left[\hat{A}{\rm Tr}
\left(2FA-\frac{i}{2}A^3\right)\right]\right\} \non
&=&-i\frac{N_c}{96\pi^2}\int d^4x \,du \left\{\frac{3}{2}\hat{A}_\mu \left(F_{\nu\rho}^aF_{\sigma\lambda}^a+\frac{1}{3}
\hat{F}_{\nu\rho}\hat{F}_{\sigma\lambda}\right) \right. \non 
&&\hspace{3cm} 
\left. +\,2\partial_\mu\left[\hat{A}_\nu\left(F_{\rho\sigma}^aA_\lambda^a+\frac{1}{4}\epsilon_{abc}A_\rho^aA_\sigma^bA_\lambda^c\right)\right]
\right\}\epsilon^{\mu\nu\rho\sigma\lambda} \, ,
\eea
where, in the first line, we have used a notation in terms of differential forms in order to connect our expression to 
the one from Ref.\ \cite{Hata:2007mb} (our integration range is $u_{\rm KK} <u<\infty$; therefore, in order to integrate over D8
and $\overline{\rm D8}$ branes we need an additional factor 2 in the prefactor compared to Eq.\ (2.8) in Ref.\ \cite{Hata:2007mb}).
The change of numerical prefactors in going from the first to the second line 
comes from performing the trace and from our convention of the field strength (\ref{Fmunu}) (the factors for the latter are hidden in the 
wedge products in the first line).

%%%%%%%%%%%%%%%%%%%%%%%%%%%%%%%%%%%%%%%%%%%%%%%%%%%%%%%%%%%%%%%%%%%%%%%%%%%%%%%%%%%%%%%
\section{Equations of motion and free energy in the chirally broken phase}
\label{confinedphase}
%%%%%%%%%%%%%%%%%%%%%%%%%%%%%%%%%%%%%%%%%%%%%%%%%%%%%%%%%%%%%%%%%%%%%%%%%%%%%%%%%%%%%%%

We can now derive the equations of motion for the gauge fields and the general form of the free energy, to be specified for various phases later.
In this section and in the entire main part of the paper, we shall be concerned with the confined, i.e., chirally broken, phase whose metric $g$ 
is given in Eq.\ (\ref{dsD8conf}). For completeness we present the equations of motion and the free energy of the deconfined, i.e., 
chirally restored, phase in Appendix \ref{sec:restored}. 

%%%%%%%%%%%%%%%%%%%%%%%%%%%%%%%%%%%%%%%%%%%%%%%%%%%%%%%%%%%%%%%%%%%%%%%%%%%%%%%%%%%%%%%%%%%%
\subsection{Equations of motion and ansatz
including magnetic field, chemical potentials, and supercurrents}
\label{sec:eqsmotion}
%%%%%%%%%%%%%%%%%%%%%%%%%%%%%%%%%%%%%%%%%%%%%%%%%%%%%%%%%%%%%%%%%%%%%%%%%%%%%%%%%%%%%%%%%%%%

We start by taking the variation with respect to the gauge fields of the Yang-Mills and Chern-Simons Lagrangians ${\cal L}_{\rm YM}$ 
and ${\cal L}_{\rm CS}$. They are given by the integrands (including the prefactors outside the integral) 
of the actions in (\ref{SYM}) and (\ref{SCS}). We present the general form of the variations in Appendix \ref{Appeqs}. Here we proceed
by using those general expressions for our specific ansatz.

The equations of motion obtained from the variations (\ref{dYM1}), (\ref{dYM2}), (\ref{dCS1}), (\ref{dCS2}) are complicated
coupled nonlinear differential equations for the gauge fields. We shall now simplify these equations by  
transforming the holographic coordinate $u$,  by choosing a particular gauge, and by choosing a specific ansatz for the fields that 
captures the physics we are interested in. The new coordinate $z$ we shall use from now on is defined through
\be \label{z}
u=(u_{\rm KK}^3+u_{\rm KK} z^2)^{1/3} \, .
\ee
We have $z\in [-\infty,\infty]$ while $u\in [u_{\rm KK},\infty]$.
In the new coordinate, the boundaries of the connected D8 and $\overline{\rm D8}$ branes correspond to $z=-\infty$ for $x_4=0$ 
(``left-handed fermions'')
and $z=+\infty$ for $x_4=L=\pi/M_{\rm KK}$ (``right-handed fermions''),  while the point $z=0$ corresponds
to the tip of the cigar-shaped $z$-$x_4$ subspace in the bulk. We work in a gauge where ${\cal A}_z=0$ \cite{Sakai:2004cn,Aharony:2007uu},
see Sec.\ \ref{sec:rotate} for a discussion of this choice.

Now we specify our ansatz for the gauge fields. First, we set all 
components proportional to $\tau_1$ and $\tau_2$ in flavor space to zero and may then, for notational convenience drop the superscript 
3 from the gauge fields and field strengths. Consequently, in the following we only have gauge fields and field strengths
with a hat ($\hat{A},\hat{F}$), corresponding to the ${\bf 1}$-components, and without any flavor index ($A,F$), corresponding to the 
$\tau_3$-components. This choice simplifies the calculations significantly but is a restriction for the possible chiral condensates 
we can capture, as we shall explain in Sec.\ \ref{sec:rotate}. 

The magnetic field is introduced as follows. The electromagnetic gauge group with generator 
$Q={\rm diag}(q_1,q_2)$, where 
$q_1$ and $q_2$ are the electric charges of the quark flavors, is a subgroup of $U(2)_L\times U(2)_R$. The magnetic
field ${\cal B}_{\rm em}$ thus has baryon and isospin components, $Q{\cal B}_{\rm em} = \hat{\cal B}{\bf 1} + {\cal B}\tau_3$, or
\be \label{Bem}
\hat{\cal B} = \frac{q_1+q_2}{2}{\cal B}_{\rm em} \, , \qquad {\cal B} = \frac{q_1-q_2}{2}{\cal B}_{\rm em} \, .
\ee
We are interested in a system of up and down flavors, i.e., $q_1=2/3\,e$, $q_2=-1/3\,e$ with $e^2=4\pi/137$ 
and $\hat{\cal B}=e{\cal B}_{\rm em}/6$, ${\cal B}=e{\cal B}_{\rm em}/2$, but 
mostly we shall derive general results, keeping $\hat{\cal B}$ and ${\cal B}$ independent of each other. 
We should recall that the gauge symmetry in the 
bulk corresponds to a global symmetry at the boundary. Therefore, there is no electromagnetic gauge symmetry at the boundary, and 
in this sense ${\cal B}_{\rm em}$ is not a dynamical magnetic field. 

We consider a spatially homogeneous magnetic field and, without loss of generality, let it point into the 3-direction. This 
requires nonzero field strengths $\hat{F}_{12}$ and $F_{12}$. We can therefore choose the ansatz
\begin{subequations} \label{ansatzb}
\bea 
\hat{A}_1({\bf x},z)=-x_2 \frac{\hat{b}(z)}{2} \, , \qquad \hat{A}_2({\bf x},z)=x_1 \frac{\hat{b}(z)}{2} \, , \\   
A_1({\bf x},z)=-x_2 \frac{b(z)}{2} \, , \qquad A_2({\bf x},z)=x_1 \frac{b(z)}{2} \, , 
\eea
\end{subequations}
such that $\hat{F}_{12}(z)=\hat{b}(z)$, $F_{12}(z)=b(z)$, and the boundary values at $z=\pm\infty$ 
of $\hat{b}(z)$, $b(z)$ given by $\hat{\cal B}$, ${\cal B}$.    
%In the case of $z$ dependent $\hat{b}(z)$, $b(z)$, we have nonzero field strengths $\hat{F}_{iz}$, $F_{iz}$. 
%They depend on ${\bf x}$ but are required to vanish at the boundary $z=\pm\infty$. 
(Note that for non-constant $\hat{b}(z)$, $b(z)$, we also have 
nonzero field strengths $\hat{F}_{iz}$, $F_{iz}$.)

Next we account for the chemical potentials. This is done by relating the boundary values at $z=\pm\infty$ for the gauge fields 
$\hat{A}_0(z)$ and $A_0(z)$ with the baryon and isospin chemical potentials  $\mu_B$ and $\mu_I$
\cite{Rozali:2007rx,Horigome:2006xu,Parnachev:2007bc}. Consequently, 
we may have nonzero field strengths $\hat{F}_{0z}$, $F_{0z}$. It turns out that within this ansatz nonzero values of the spatial gauge fields
may be induced, i.e., we have to take into account $\hat{A}_3(z)$, $A_3(z)$ and thus the field strengths $\hat{F}_{3z}$, $F_{3z}$.
The boundary values at $z=\pm\infty$ of the spatial gauge fields are identified with the gradients of the meson fields 
\cite{Sakai:2004cn,Bergman:2008qv}. These gradients correspond, according to the usual hydrodynamic theory of a superfluid 
\cite{vollhardt,Son:2002zn}, to ``supercurrents'', i.e., currents of the condensate, in our context for instance the current of a pion condensate;
see also Refs.\ \cite{Kryjevski:2008zz,Schafer:2005ym}.   
Consequently, we shall identify $\hat{A}_3(\pm\infty)$, $A_3(\pm\infty)$ with meson supercurrents $\hat{\jmath}$, $\jmath$. 
The supercurrents are not external parameters, hence we shall minimize
the free energy with respect to them \cite{Bergman:2008qv,Thompson:2008qw}. They should not be confused with the ``normal'' 
currents $\mathcal J_i=\delta S_{\rm eff}/\delta A^i$, discussed in the Sakai-Sugimoto model in detail in
Refs.\ \cite{Hata:2008xc,Hashimoto:2008zw}, and computed below in Sec.\ \ref{sec:sigma}. 
The supercurrents rather act as a source for the normal currents. 

Now we can insert the ansatz and the coordinate transformation (\ref{z}) into the general equations of motion 
(\ref{dYM1}), (\ref{dYM2}), (\ref{dCS1}), (\ref{dCS2}). We then need to replace ${\cal A}_0\to i{\cal A}_0$, 
since we are working in euclidean space.
We find the following equations for the magnetic field 
\be \label{magn}
\partial_z[k(z)\partial_z \hat{b}]=\partial_z[k(z)\partial_z b]=0 \, , 
\ee
where
\be \label{k}
k(z)\equiv u_{\rm KK}^3+u_{\rm KK} z^2 \, .
\ee
They arise from the Yang-Mills variation with respect to the spatial gauge field, Eqs.\ (\ref{dYM12}) and (\ref{dCS12}) 
and contain no contribution from the Chern-Simons term. Moreover, they decouple from the equations for the other fields,
which are 
\begin{subequations} \label{diffeqs}
\bea
\partial_z[k(z)\hat{F}_{z0}] &=& \frac{\alpha u_{\rm KK}^2}{M_{\rm KK}^2}\left[b(z)F_{z3}+\hat{b}(z)\hat{F}_{z3}\right]\, , \\
\partial_z[k(z)F_{z0}] &=& \frac{\alpha u_{\rm KK}^2}{M_{\rm KK}^2}\left[b(z)\hat{F}_{z3}+\hat{b}(z)F_{z3}\right] \, , \\
\partial_z[k(z)\hat{F}_{z3}] &=& \frac{\alpha u_{\rm KK}^2}{M_{\rm KK}^2}\left[b(z)F_{z0}+ \hat{b}(z)\hat{F}_{z0}\right]\, , \\ 
\partial_z[k(z)F_{z3}] &=& \frac{\alpha u_{\rm KK}^2}{M_{\rm KK}^2}\left[b(z)\hat{F}_{z0}+\hat{b}(z)F_{z0}\right]\, ,
\eea
\end{subequations}
where
\be \label{alpha}
\alpha\equiv \frac{27\pi}{2\lambda} \, .
\ee
In all four equations in (\ref{diffeqs}) the left-hand side comes from the variation of the Yang-Mills contribution, while the right-hand 
side originates from the Chern-Simons contribution. We see that the latter is proportional to the magnetic field. 
The equations (\ref{magn}) and (\ref{diffeqs}) shall be solved analytically in Sec.\ \ref{sec:broken}. Before doing so we use these equations to 
derive a simple expression for the free energy.

%%%%%%%%%%%%%%%%%%%%%%%%%%%%%%%%%%%%%%%%%%%%%%%%%%%%%%%%%%%%%%%%%%%%%%%%%%%%%%%%%%%%%%%%%%%%
\subsection{Free energy and holographic renormalization}
\label{sec:free}
%%%%%%%%%%%%%%%%%%%%%%%%%%%%%%%%%%%%%%%%%%%%%%%%%%%%%%%%%%%%%%%%%%%%%%%%%%%%%%%%%%%%%%%%%%%%

With the metric of the confined phase (\ref{dsconf}), 
the relations between the parameters of the model (\ref{R3}), (\ref{MKK}), (\ref{coupling}), and 
the new coordinate $z$ (\ref{z}), the Yang-Mills part of the action (\ref{SYM}) can be written as
\be \label{YM5}
S_{\rm YM} = \kappa\int d^4x \int_{-\infty}^\infty dz\, \left\{\frac{16M_{\rm KK}^2k^{2/3}(z)}{9(2\pi\alpha')^2u_{\rm KK}} +
\frac{M_{\rm KK}^2}{u_{\rm KK}^2}k(z){\rm Tr}[{\cal F}_{z\mu}^2]+\frac{1}{2}h(z)
{\rm Tr}[{\cal F}_{\mu\nu}^2]\right\} \, , 
\ee
where $\mu,\nu=0,1,2,3$. Here, $k(z)$ is given in Eq.\ (\ref{k}),  
\be
h(z) \equiv (u_{\rm KK}^3+u_{\rm KK} z^2)^{-1/3} \, ,
\ee
and 
\be \label{kappa}
\kappa\equiv\frac{\lambda N_c}{216\pi^3} \, .
\ee
In deriving Eq.\ (\ref{YM5}) we have used that the field strengths are symmetric or antisymmetric functions of $z$. 
We shall see later that this is indeed the 
case for all phases we consider. This form of the action is general, and it is straightforward to insert our
ansatz discussed in the previous subsection. 

To compute the Chern-Simons contribution to the free energy we first note
that the surface term (last term on the right-hand side of Eq.\ (\ref{SCS})) gives a nonzero contribution. Within our ansatz the term 
$\propto d(\hat{A}{\rm Tr}[A^3])$ vanishes since our
only nonzero flavor components of the gauge fields are proportional to ${\bf 1}$ and $\tau_3$; however, the term 
$\propto d(\hat{A}{\rm Tr}[FA])$ does not vanish. We find 
\begin{subequations}
\bea
\hat{A}_\mu F_{\nu\rho}F_{\sigma\lambda}\epsilon^{\mu\nu\rho\sigma\lambda} &=& 8b(\hat{A}_3F_{z0}-\hat{A}_0F_{z3}) \, , \\
\hat{A}_\mu\hat{F}_{\nu\rho}\hat{F}_{\sigma\lambda}\epsilon^{\mu\nu\rho\sigma\lambda} 
&=& 8\hat{b}(\hat{A}_3\hat{F}_{z0}-\hat{A}_0\hat{F}_{z3}) \, , \\
\partial_\mu(\hat{A}_\nu F_{\rho\sigma}A_\lambda)\epsilon^{\mu\nu\rho\sigma\lambda} &=& 
2b(A_3\hat{F}_{z0}-A_0\hat{F}_{z3} +2\hat{A}_0F_{z3}-2\hat{A}_3F_{z0}) \non
&&+\,2\hat{b}(A_3F_{z0}-A_0F_{z3}) \, . \label{surface}
\eea
\end{subequations}
Inserting these expressions into the Chern-Simons action (\ref{SCS}) yields, with ${\cal A}_0\to i {\cal A}_0$ and 
$N_c/(16\pi^2)=\alpha\kappa$, 
\bea
S_{\rm CS}&=& \frac{\alpha\kappa}{3}\int dx^4 \int_{-\infty}^\infty dz\,\left[ \hat{b}\left(\hat{A}_3\hat{F}_{z0}+A_3F_{z0}-
\hat{A}_0\hat{F}_{z3}-A_0F_{z3}\right) \right. \non
&& \left. \hspace{3cm} +\, b\left(\hat{A}_3F_{z0}+A_3\hat{F}_{z0}-
\hat{A}_0F_{z3}-A_0\hat{F}_{z3}\right)\right] \non
&=&\frac{\kappa M_{\rm KK}^2}{3u_{\rm KK}^2}\frac{V}{T}\left\{\int_{-\infty}^\infty dz\,k(z)(\hat{F}_{z0}^2+F_{z0}^2-\hat{F}_{z3}^2-F_{z3}^2) \right.
\non
&& \left. \hspace{2cm} -\,\left[k(z)(\hat{A}_0\hat{F}_{z0}+A_0F_{z0}-\hat{A}_3\hat{F}_{z3}-A_3F_{z3})\right]_{z=-\infty}^{z=+\infty}\right\} \, ,
\eea
where, in the second step, we have used the equations of motion (\ref{diffeqs}), 
and where $V$ is the three-dimensional volume of space and $T$ the temperature. In changing the integration over the holographic 
coordinate from $u\in[u_{\rm KK},\infty]$ to $z\in[-\infty,\infty]$ we have assumed that the integrand is symmetric in $z$. In all phases
we consider this turns out to be the case. 
Putting the Yang-Mills and Chern-Simons contribution together, we obtain the free energy density $\Omega \equiv T(S_{\rm YM}+S_{\rm CS})/V$ 
\bea \label{Omegaconf}
\Omega &=& 
\Omega_g+ \Omega_b + \frac{\kappa M_{\rm KK}^2}{6u_{\rm KK}^2}\int_{-\infty}^\infty dz\,k(z)(-\hat{F}_{z0}^2-F_{z0}^2+\hat{F}_{z3}^2+F_{z3}^2)
\non 
&&-\,\frac{\kappa M_{\rm KK}^2}{3u_{\rm KK}^2}\left[k(z)(\hat{A}_0\hat{F}_{z0}+A_0F_{z0}-\hat{A}_3\hat{F}_{z3}-A_3F_{z3})
\right]_{z=-\infty}^{z=+\infty}  \, ,
\eea
where the geometric contribution $\Omega_g$ is given by the field-independent first term on the right-hand side of Eq.\ (\ref{YM5}). This term is 
independent of all gauge fields and field strengths and thus plays no role in 
discussing the physical properties of a given phase. 
%For free energy comparisons of phases with different geometries one can handle this term 
%with the proper renormalization explained in Ref.\ \cite{Witten:1998zw}. 
Moreover, we shall only compare free energies of phases with  
identical embedding of the flavor branes. Hence, for our purpose, this term can simply be dropped from now on.
The term $\Omega_b$ in Eq.\ (\ref{Omegaconf}) is given by
\bea \label{OmegaH}
\Omega_b &\equiv& \frac{\kappa}{2} \int_{-\infty}^{\infty} dz\,h(z)[\hat{b}^2(z)+b^2(z)] \non
&& + \, \frac{\kappa M_{\rm KK}^2}{4u_{\rm KK}^2}\frac{\int dx_1 dx_2(x_1^2+x^2_2)}{\int dx_1 dx_2} \int_{-\infty}^{\infty}
dz \, k(z)[(\partial_z\hat{b})^2+(\partial_z b)^2] \, .
\eea
Both contributions of $\Omega_b$ solely depend on the magnetic field (remember that the equations of motion for $\hat{b}$ and $b$ (\ref{magn}) decouple 
from the other field equations). Therefore, $\Omega_b$ is irrelevant for minimizing the free energy with respect to the supercurrents 
$\hat{\jmath}$ and $\jmath$. However, it can play a role when
comparing free energies. This poses a problem, as
both terms of $\Omega_b$ are divergent. 

Let us first consider only constant functions 
$b(z)={\cal B}$ and $\hat b(z)=\hat{\cal B}$, for which
only the first term in $\Omega_b$ is present. Since we have already divided
by the volume $V$ of 3-space, we would expect a finite energy density from
a homogeneous magnetic field, but because of the extra holographic dimension,
this is not the case. In fact, since $h(z)\sim z^{-2/3}$, the divergence of
$\Omega_b$ comes from the $|z|\to\infty$ limits of integration and
is thus a 
%The origin of the first term in $\Omega_b$ is easy to understand: a 
%constant, space-filling magnetic field should yield a contribution to the energy which is proportional to the volume. Since we have already divided 
%out the three-dimensional spatial volume in the free energy density $\Omega_b$ we are left with the integration over the bulk, yielding 
%an infinite energy density. This is a 
typical holographic divergence 
which can be treated by holographic renormalization
\cite{Karch:2005ms}. Here we do not attempt to provide a complete discussion of this procedure, which
for the (nonconformal) Sakai-Sugimoto model has been introduced only recently \cite{Wiseman:2008qa,Kanitscheider:2008kd}. 
We rather follow the method outlined in these papers and subtract a counterterm, fixed by a physical renormalization condition,
as follows. After restricting the holographic integration in $\Omega_b$ to 
a finite interval $-\Lambda < z < \Lambda$, we subtract a counterterm $\delta\Omega_b(\Lambda)$ which cancels the divergence and obtain
a renormalized contribution $\Omega_b^{\rm ren}$. We also include a finite
counterterm which is fixed
by requiring the free energy in the absence of any chemical potential to 
vanish,
\be \label{Omegavanish}
\Omega(\mu_{B,I}=0) = 0 \, . 
\ee
This condition is motivated by the observation that $\Omega$ should be the matter part of the free energy, i.e., it should describe
the fermions and their interaction with the magnetic field. 
In particular, we thus require that the energy density of the
(nondynamical) magnetic field in the absence of any matter be left out.
%It should not yield the field energy which should rather 
%be added externally. 
This we shall later treat separately
when we consider the Gibbs free energy (the Legendre transform
 from fixed internal magnetic field to fixed external magnetic field)
in Sec.\ \ref{sec:compare}.
The condition (\ref{Omegavanish}) implies that we have to require 
\be \label{renorm}
0=\Omega_b^{\rm ren} \equiv \lim_{\Lambda\to\infty} [\Omega_b(\Lambda)-\delta\Omega_b(\Lambda)] \, .
\ee
To find the exact form of the counterterms we first note that, for
constant $\hat{b}(z)=\hat{\cal B}$ and  $b(z)={\cal B}$,
\be \label{ObL}
\Omega_b(\Lambda) = 3\kappa(\hat{\cal B}^2+{\cal B}^2)\left[\frac{\Lambda^{1/3}}{u_{\rm KK}^{1/3}} 
-\sqrt{\pi}\,\frac{\Gamma(5/6)}{\Gamma(1/3)}+{\cal O}\left(\frac{u_{\rm KK}^{5/3}}{\Lambda^{5/3}}\right)\right]  \, .
\ee
%where we have assumed that $\hat{b}(z)=\hat{\cal B}$ and  $b(z)={\cal B}$ are constant. 
%This will turn out to be the case for all phases we consider. Furthermore, we have
The counterterm $\delta\Omega_b(\Lambda)$ 
should depend only on fields and geometric data on the slice 
$z=\Lambda$, in particular it should only involve the induced metric $\gamma_{\mu\nu}$ on the slice and not the complete metric $g$.
%has the same geometric form as the original term, but, instead of the metric $g$,  
%involves the metric $\gamma$ at the slice 
%$z=\Lambda$ (i.e., $\gamma$ is obtained from 
%the original metric (\ref{dsD8conf}) by dropping the term proportional to $du^2$, then transforming the $u$ coordinate into $z$, and then setting 
%$z=\Lambda$). 
By including appropriate factors of the dilaton \cite{Wiseman:2008qa,Kanitscheider:2008kd} and an appropriate numerical 
factor to fulfill the condition (\ref{renorm})
we find
\be \label{dObL}
\delta\Omega_b(\Lambda) = \frac{R}{2}\left[\left(\frac{e^\Phi}{g_s}\right)^{1/3}-\frac{\sqrt{\pi}\,\Gamma(5/6)}{\Gamma(1/3)}
\left(\frac{u_{\rm KK}}{R}\right)^{1/2}\left(\frac{e^\Phi}{g_s}\right)^{-1/3}\right] {\cal C}(\Lambda)\, , 
\ee
with
\bea
{\cal C}(\Lambda)&\equiv& 
-\frac{T_8V_4(2\pi\alpha')^2}{2} e^{-\Phi}\sqrt{\gamma}\,\gamma^{\mu\nu}\gamma^{\rho\sigma}{\rm Tr}[{\cal F}_{\nu\rho}{\cal F}_{\sigma\mu}] 
\non
&=&\frac{6\kappa}{u_{\rm KK}^{1/4}R^{3/4}}\left[\frac{\Lambda^{1/6}}{u_{\rm KK}^{1/6}}+{\cal O}\left(\frac{1}{\Lambda^{11/6}}\right)\right] \, , 
\eea
where we have used Eq.\ (\ref{help3}) and where the indices $\mu,\nu,\rho,\sigma$ run over $0,1,2,3$. 
With this counterterm,
the term proportional to $\Lambda^{1/3}$ ($\Lambda^0$) in Eq.\ (\ref{ObL})
is cancelled by the first (second) term in Eq.\ (\ref{dObL})

In the case of a magnetic field which is not constant in the bulk,
the second term in $\Omega_b$ as given by Eq.\ (\ref{OmegaH})
is also divergent, but its divergence comes from the integration
over the spatial directions perpendicular to the magnetic field,
regardless of whether the holographic $z$-integration is finite or not.
%The second term, which is nonvanishing in the case of a 
%magnetic field nonconstant in $z$, seems curious. A nonconstant field seems to induce an infrared divergence, depending on the spatial directions
%transverse to the magnetic field. This appears to be unphysical since $\Omega_b$ is an energy {\it density} which has to be finite. 
%Note that the holographic $z$-integration may well be finite, and yet this term is divergent for an infinite three-dimensional space volume.
Therefore, we cannot treat this term by the usual holographic renormalization
and we interpret this divergence, when present, as a Meissner effect:
a phase where a homogeneous magnetic field $\mathcal B_{\rm em}$,
which fixes the boundary values of 
$\hat{b}(z)$ and $b(z)$,
is only possible for non-constant functions in $z$, is infinitely
penalized such that only $\mathcal B_{\rm em}=0$ is allowed.
As we shall see, this will be the case for the charged pion condensate,
to be discussed further in 
Sec.\ \ref{sec:meissner}. At this point we already observe that the role of the spatial directions transverse to the magnetic
field is no coincidence. It points to the necessity of currents in these directions which produce a magnetic field equal in magnitude but with 
opposite direction compared to the 
external magnetic field. This leads to a vanishing total magnetic field in the system, which is nothing but the Meissner effect for superconductors.

%%%%%%%%%%%%%%%%%%%%%%%%%%%%%%%%%%%%%%%%%%%%%%%%%%%%%%%%%%%%%%%%%%%%%%%%%%%%%%%%%%%%%%%%%%%%
\section{Chirally broken phases in a magnetic field}
\label{sec:broken}
%%%%%%%%%%%%%%%%%%%%%%%%%%%%%%%%%%%%%%%%%%%%%%%%%%%%%%%%%%%%%%%%%%%%%%%%%%%%%%%%%%%%%%%%%%%%

In this section we solve the equations of motion for the chirally broken phase.
We shall distinguish between two different chirally broken phases, the $\sigma$ and the $\pi$ phase. 
This is the main part of the paper, and the main physical results can be found in Secs.\ \ref{sec:properties1}
and \ref{sec:compare}.

%%%%%%%%%%%%%%%%%%%%%%%%%%%%%%%%%%%%%%%%%%%%%%%%%%%%%%%%%%%%%%%%%%%%%%%%%%%%%%%%%%%%%%%%%%%%
\subsection{Chiral rotations and resulting boundary conditions}
\label{sec:rotate}
%%%%%%%%%%%%%%%%%%%%%%%%%%%%%%%%%%%%%%%%%%%%%%%%%%%%%%%%%%%%%%%%%%%%%%%%%%%%%%%%%%%%%%%%%%%%

In $N_f=2$ chiral perturbation theory the chiral field $U\in U(2)$ describing the Goldstone bosons is given by 
\be \label{chiral}
U=e^{i (\eta+\varphi_a\tau_a)/f_\pi} \, , 
\ee
where $f_\pi$ is the pion decay constant (in the Sakai-Sugimoto model, $f_\pi=2M_{\rm KK}\sqrt{\kappa/\pi}$ \cite{Sakai:2004cn,Hashimoto:2008zw}). 
The $\eta$ meson (the $\eta'$ for $N_f=3$)
becomes massive in QCD due to the explicit breaking of the $U(1)_A$
through the axial anomaly. 
This is realized in the Sakai-Sugimoto model through the 
Chern-Simons term, and the mass of the $\eta$ can be computed within the model, 
$m_{\eta} = \lambda M_{\rm KK} \sqrt{N_f/N_c}/(3\sqrt{3}\pi)$ \cite{Sakai:2004cn}, see also 
Refs.\ \cite{Armoni:2004dc,Barbon:2004dq,Bergman:2006xn,Parnachev:2008fy}. 
%For the purpose of the following arguments 
%By implicitly taking into account a $\theta$ term,
%we may use the full $U(2)_L\times U(2)_R$ 
%symmetry. %We comment on this assumption below Eq.\ (\ref{gradeta}).

In the Sakai-Sugimoto model the chiral field is given by the holonomy \cite{Sakai:2004cn}
\be 
U=P\exp\left(i\int_{-\infty}^\infty dz\,{\cal A}_z\right) \, .
\ee
As mentioned above, we work in a gauge where ${\cal A}_z=0$. 
This is only possible by using the full $U(2)_L\times U(2)_R$ symmetry,
implicitly taking into account a $\theta$ term, see also comment
below Eq.\ (\ref{gradeta}).
It seems we can then only consider the vacuum $U={\bf 1}$. 
However, we can keep the ${\cal A}_z=0$ gauge and recover other vacua
encoded in the boundary values of the gauge fields. This is explained in detail for instance in Ref.\ \cite{Aharony:2007uu}. We shall now 
recapitulate this explanation and apply it to our case.

Consider a potential $V[\mu_L,\mu_R,U(\phi)]$ which is invariant under 
$U(N_f)_L\times U(N_f)_R$.  Here, $\mu_L, \mu_R \in U(N_f)$ are fixed external parameters. For the following argument we denote these parameters
simply by $\mu_L$, $\mu_R$, reminiscent of the chemical potentials, but one should keep in mind that this notation also includes the  
magnetic field.
The chiral field $U$ is written as a function of a parameter $\phi$ with respect to which we have to minimize
the potential to find the vacuum. This parameter is a symbol for the meson fields in Eq.\ (\ref{chiral}). 
The external parameters transform under the global symmetry as $\mu_L\to g_L^{-1}\mu_L g_L$, $\mu_R\to g_R^{-1}\mu_R g_R$, while 
the chiral field transforms as $U\to g_L^{-1} U g_R$, where $g_L\in U(N_f)_L$, $g_R\in U(N_f)_R$. 
Via a global symmetry transformation we have 
$V[\mu_L,\mu_R,U(\phi)]=V[g_L^{-1}(\phi)\mu_Lg_L(\phi),g_R^{-1}(\phi)\mu_Rg_R(\phi),{\bf 1}]$ with 
$\phi$-dependent transformations $g_L(\phi)$, $g_R(\phi)$ such that $g_L^{-1}(\phi) U(\phi)  g_R(\phi) = {\bf 1}$. 
To find the vacuum of the theory it obviously does not matter whether we use the original potential or the potential with the transformed
quantities because both expressions are simply identical.
Consequently, instead of keeping the external parameters fixed and varying the chiral field we can fix the chiral field to be the 
unit matrix and vary the external parameters. Of course we cannot simply treat the external parameters as arbitrary continuous quantities
with respect to which we minimize the potential. We need to ensure that they are connected by a transformation to their physical values. 
We shall see below that within our ansatz the allowed rotated parameters only assume two discrete values, such that we
simply have to compare two separate phases with each other. 
After minimization of the potential, the physical vacuum is given by applying the rotation found from 
minimization ``backwards'' onto the unit matrix, i.e., 
\be \label{gRgL}
U=g_Lg_R^{-1} \, .
\ee
Without loss of generality we can set $g_R={\bf 1}$ and thus $U=g_L$. We can write 
\be \label{g}
g_L=e^{i(\eta+ \varphi_a\tau_a)/f_\pi} =e^{i\eta/f_\pi}\frac{\sigma + i \pi_a\tau_a}{f_\pi} \, , 
\ee
where $\sigma/f_\pi\equiv \cos(\varphi/f_\pi)$, $\pi_a/f_\pi=\varphi_a/\varphi\,\sin(\varphi/f_\pi)$ with 
$\varphi\equiv (\varphi_1^2+\varphi_2^2+\varphi_3^2)^{1/2}$. This is the
usual form of the chiral field in chiral perturbation theory, where the massive mode, the ``sigma'', is frozen and the effective theory
describes the remaining meson modes. 
Therefore, both sides of Eq.\ (\ref{g}) contain 
four degrees of freedom; for the right-hand
side we have the condition $\sigma^2+\pi^2=f_\pi^2$ which is obvious from the definitions of $\sigma$ and $\pi_a$.

To apply a rotation given by $g_L$ on the (left-handed) external parameter $\mu_L$ note that 
our physical chemical potentials and the magnetic field are diagonal in flavor space and identical for $L$ and $R$,
$\mu_L=\mu_R=\mu_B{\bf 1}+\mu_I\tau_3$, ${\cal B}_{{\rm em},L}={\cal B}_{{\rm em},R}=\hat{\cal B}{\bf 1}+{\cal B}\tau_3$.
The baryon part $\propto {\bf 1}$ does obviously not change under a $U(2)$ transformation. We thus only have to consider how the 
isospin part $\propto \tau_3$ transforms. We find
\bea \label{transform}
g^{-1}_L\tau_3 g_L &=& \frac{1}{f_\pi}\left[\pi^+(\pi^0+i\sigma)\tau_+ + \pi^-(\pi^0-i\sigma)\tau_-+(1-2\pi^+\pi^-)\tau_3\right]\, , 
\eea
where $\tau_\pm\equiv \tau_1\pm i\tau_2$, and where we have introduced the neutral pion $\pi^0=\pi_3$ and the charged pions 
$\pi^\pm\equiv \pi_1\mp i\pi_2$. 
In our ansatz described in Sec.\ \ref{sec:eqsmotion} we have restricted ourselves to diagonal gauge fields. Since the chemical potentials and 
the magnetic field are the boundary values for the gauge fields, they have to be diagonal too. Consequently, we can only 
allow for transformations (\ref{transform}) that transform $\tau_3$ into a matrix $\propto\tau_3$. There are two (nontrivial) 
possibilities to make the coefficients in front of $\tau_+$, $\tau_-$ vanish: $(i)$ $\pi^+=\pi^-=0$ which leads to 
$g^{-1}_L\tau_3 g_L= \tau_3$ and $(ii)$ $\pi^0=\sigma=0$ which leads to $g^{-1}_L\tau_3 g_L= -\tau_3$. Hence we can either 
leave the isospin components of the chemical potentials and the magnetic field invariant or flip their sign. This means that 
the parameter $\phi$ in $U(\phi)$ above is in fact discrete, not continuous. Had we allowed for off-diagonal components 
in the gauge fields, we could have described arbitrary linear combinations of the pion fields. 

\TABLE[t]{
\begin{tabular}{|c||c|c|c|c|c|c|} 
\hline
\rule[-1.5ex]{0em}{4ex} & $\;\;\hat{A}_0(\pm\infty)\;\;$ & $\;\;A_0(\pm\infty)\;\;$ & $\;\;\hat{b}(\pm\infty)\;\;$ & $\;\;b(\pm\infty)\;\;$ &
$\;\;\hat{A}_3(\pm\infty)\;\;$ & $\;\;A_3(\pm\infty)\;\;$ 
 \\ \hline\hline
\rule[-1.5ex]{0em}{4ex}$\;\;\sigma\;\;$ & $2\mu_B$ & $2\mu_I$ & $\hat{\cal B}$ & ${\cal B}$ & $\pm 2\hat{\jmath}$ & $\pm 2\jmath$  \\ \hline
\rule[-1.5ex]{0em}{4ex}$\;\;\pi\;\;$ & $2\mu_B$ & $\pm 2\mu_I$ & $\hat{\cal B}\; (0) $ & $\pm {\cal B}\; (0) $ & $\pm 2\hat{\jmath}$ & 0   
\\ \hline
\end{tabular}
\caption{Boundary conditions in the sigma and pion phases. The boundary conditions for the temporal components of 
the gauge fields correspond to the baryon and isospin chemical potentials, while 
the boundary conditions for the field strengths $\hat{F}_{12}\equiv \hat{b}$, $F_{12}\equiv b$ correspond to the baryon and isospin 
components of the magnetic field. The boundary conditions for the spatial components 
$\hat{A}_3$, $A_3$ are given by the meson supercurrents $\hat{\jmath}$, $\jmath$. These currents are not external parameters 
but have to be determined by minimizing the free energy. In the $\pi$ phase $A_3(\pm\infty)$ has to vanish to ensure a 
well-defined behavior of the gauge fields under parity transformations, see Eq.\ (\ref{JIjI}) and discussion above this equation.
For a discussion of the normalization of the chemical
potentials see Eq. (\ref{J0dOdmu}) and below.
The zeros in parantheses for the magnetic fields in the charged pion 
condensed phase indicate that eventually we shall set $\hat{\cal B}={\cal B}=0$ because of the Meissner effect in this phase, see
Sec.\ \ref{sec:meissner}. 
}
\label{tableboundary}
}

These somewhat formal arguments have a very intuitive geometric interpretation \cite{Parnachev:2007bc}: another (simpler, but less precise) 
way of saying what we have just 
explained is the following. Think of the D8 branes as a left-handed up-brane and a left-handed down-brane and of the 
$\overline{\rm D8}$ branes as a right-handed up-brane and a right-handed down-brane. Then, a chirally broken phase can be 
constructed by connecting $(i)$ the left-handed up-brane with the right-handed up-brane and likewise for the down-branes or $(ii)$ the 
left-handed up-brane with the right-handed down-brane and vice versa.
These two possibilities correspond exactly to the two cases from the above formal argument: case $(i)$ corresponds to a condensate 
where equal quark flavors participate, i.e., a combination of $\sigma$ and $\pi^0$ with  
$\bar{u}-u$ and $\bar{d}-d$ pairing. 
In the remainder of the paper we shall refer to this case as the 
$\sigma$ phase. Case $(ii)$ corresponds to a charged pion condensate with nonzero  
$\langle\bar{d}\gamma_5 u\rangle$, $\langle\bar{u}\gamma_5 d\rangle$, to which we shall refer
as the $\pi$ phase (this phase is sometimes called ``$\rho$'' \cite{Parnachev:2007bc,Klein:2003fy}). 
As a summary of this section and a reminder for the subsequent sections,  we present the resulting boundary conditions for the 
$\sigma$ and the $\pi$ phases in Table \ref{tableboundary}. 

In this Table we have also included the supercurrents $\hat{\jmath}$, $\jmath$, which, in our gauge $g_L=U$, have the form $g_L^{-1}\nabla g_L$ 
\cite{Sakai:2004cn}. With Eq.\ (\ref{g}) this becomes for the two phases
\begin{subequations} \label{gradeta}
\bea
&(i)\;\;\sigma \;{\rm phase}:& \qquad -ig_L^{-1}\nabla g_L = \frac{\nabla\eta}{f_\pi}+\frac{\tau_3}{f_\pi^2}
\left(\sigma\nabla\pi^0-\pi^0\nabla\sigma\right) \, , \\[2ex]
&(ii)\;\;\pi \;{\rm phase}:& \qquad -ig_L^{-1}\nabla g_L = \frac{\nabla\eta}{f_\pi}+\frac{i\tau_3}{2f_\pi^2}
\left(\pi^-\nabla\pi^+-\pi^+\nabla\pi^-\right) \, . \qquad
\eea
\end{subequations}
We see that the supercurrents are diagonal, i.e., our ansatz with 
nonvanishing ${\bf 1}$-component $\hat{\jmath}$ and $\tau_3$-component $\jmath$ is consistent. 
Interestingly, an anisotropic $\eta$ condensate appears in the ${\bf 1}$-components $\hat{\jmath}$. 
The $\eta$ condensate has dropped out in Eq.\ (\ref{transform}), and thus our boundary conditions, given by the 
chemical potentials and the magnetic field modified by the rotation (\ref{transform}), do not reveal whether 
there is an admixture of an $\eta$ condensate in the $\sigma$ phase. On the other hand, a nonzero supercurrent $\hat{\jmath}$ seems to indicate 
the presence of an $\eta$ supercurrent. Indeed, we shall see later that in the $\sigma$ phase a nonzero $\hat{\jmath}$ is induced.  
The term $\nabla\eta$ in Eqs.\ (\ref{gradeta}) appears due to our use of the full $U(2)_L\times U(2)_R$ symmetry. Strictly speaking, 
our Lagrangian breaks the axial $U(1)_A$ because of the presence of the Chern-Simons term. However, this symmetry is preserved if one compensates
a $U(1)_A$ rotation by a shift of the $\theta$ parameter, whose realization in the Sakai-Sugimoto model is discussed in Ref.\ \cite{Sakai:2004cn,Bergman:2006xn,Parnachev:2008fy},
see also Ref.\ \cite{Witten:1998uka}. We thus implicitly adjust the $\theta$ parameter when using the full gauge symmetry, absorbing a constant $\eta$ mode into $\theta$. We shall proceed
within this simplification, but have to keep in mind that in a more complete approach one would have to consider a fixed $\theta$ and allow for
a constant $\eta$ mode explicitly. Such 
an approach would be of interest especially in view of recent studies of possible (CP-violating) $\eta$ condensates in an NJL model calculation 
\cite{Boer:2008ct}, or, including a magnetic field, in the linear sigma model
\cite{Mizher:2008hf}. 

We finally remark that our setup does not include the possibility 
of diquark condensation of the form $\langle ud\rangle$, 
which is expected to lead to color superconductivity of quark matter at sufficiently large baryon 
chemical potential \cite{Alford:2007xm}. However, color superconductivity does not necessarily occur in the large $N_c$ limit where 
a ``chiral density wave'' is a strong candidate for the ground state \cite{Deryagin:1992rw,Shuster:1999tn}, or, as suggested recently, quark matter 
may be confined even for large chemical potentials \cite{McLerran:2007qj,Glozman:2007tv}.  

%%%%%%%%%%%%%%%%%%%%%%%%%%%%%%%%%%%%%%%%%%%%%%%%%%%%%%%%%%%%%%%%%%%%%%%%%%%%%%%%%%%%%%%%%%%%
\subsection{Solutions of the equations of motion and free energies}
\label{sec:solve}
%%%%%%%%%%%%%%%%%%%%%%%%%%%%%%%%%%%%%%%%%%%%%%%%%%%%%%%%%%%%%%%%%%%%%%%%%%%%%%%%%%%%%%%%%%%%

We can now solve the equations of motion (\ref{magn}) and (\ref{diffeqs}) for the two sets of boundary conditions given in Table 
\ref{tableboundary}. For notational convenience we set $u_{\rm KK}=1$ (the final results in Secs.\ \ref{sec:properties1} and 
\ref{sec:compare} do not depend on $u_{\rm KK}$ and thus all physical quantities will have the correct dimensions). 
For both sets of boundary conditions we first note that the differential equations (\ref{diffeqs}) can be solved by defining the new functions
\be \label{F03}
F_0^\pm(z)\equiv k(z)\frac{\hat{F}_{z0}\pm F_{z0}}{2} \, , \qquad 
F_3^\pm(z)\equiv k(z)\frac{\hat{F}_{z3}\pm F_{z3}}{2} \, .
\ee
Then, the four equations (\ref{diffeqs}) are equivalent to 
\be \label{newdiffs}
\partial_z F_0^\pm=\frac{\alpha[\hat{b}(z)\pm b(z)]}{k(z)M_{\rm KK}^2}\,F_3^\pm(z) \, , \qquad 
\partial_z F_3^\pm=\frac{\alpha[\hat{b}(z)\pm b(z)]}{k(z)M_{\rm KK}^2}\,F_0^\pm(z) \, . 
\ee
Now the two equations with the upper sign are decoupled from the two equations with the lower sign. To proceed, we have to distinguish between
the two chirally broken phases.

%%%%%%%%%%%%%%%%%%%%%%%%%%%%%%%%%%%%%%%%%%%%%%%%%%%%%%%%%%%%%%%%%%%%%%%%%%%%%%%%%%%%%%%%%%%%
\subsubsection{Sigma phase}
\label{sec:sigma}
%%%%%%%%%%%%%%%%%%%%%%%%%%%%%%%%%%%%%%%%%%%%%%%%%%%%%%%%%%%%%%%%%%%%%%%%%%%%%%%%%%%%%%%%%%%%

With the boundary conditions of the $\sigma$ phase from Table \ref{tableboundary} and with Eqs.\ (\ref{magn}) we conclude that 
the magnetic fields are constant in the bulk 
\bea
\hat{b}(z) = \hat{\cal B} \, , \qquad b(z)={\cal B} \, . \label{notwist} 
\eea
In the following, we shall denote the dimensionless magnetic fields by 
\be \label{dimless}
\hat{B}\equiv \frac{\alpha\hat{\cal B}}{M_{\rm KK}^2}  \, , \qquad B\equiv \frac{\alpha {\cal B}}{M_{\rm KK}^2}  \, , 
\qquad B_{\rm em}\equiv \frac{\alpha {\cal B_{\rm em}}}{M_{\rm KK}^2}\, . 
\ee
We can now solve Eqs.\ (\ref{newdiffs}) for completely general boundary conditions for the gauge fields. This 
is done in Appendix \ref{appsigma}, where we present some technical details. 
Here we proceed with the specific solution obtained from the boundary conditions given in the first row of Table \ref{tableboundary}.
This solution yields the gauge fields  
\begin{subequations} \label{gaugefields1}
\bea
\hat{A}_0(z) &=& 2\mu_B + \hat{\jmath}[C_+(z)+C_-(z)-T_+]+\jmath[C_+(z)-C_-(z)-T_-] \, , \\
A_0(z) &=& 2\mu_I +\jmath[C_+(z)+C_-(z)-T_+] +\hat{\jmath}[C_+(z)-C_-(z)-T_-] \, , \\
\hat{A}_3(z) &=& \hat{\jmath}[S_+(z)+S_-(z)]+\jmath[S_+(z)-S_-(z)]  \, , \\
A_3(z) &=& \jmath[S_+(z)+S_-(z)]+ \hat{\jmath}[S_+(z)-S_-(z)] \, , 
\eea
\end{subequations}
where we have abbreviated 
\begin{subequations} \label{CST}
\bea
C_\pm(z)&\equiv& \frac{\cosh[(\hat{B}\pm B)\arctan z]}{\sinh[\pi(\hat{B}\pm B)/2]} \, , \qquad 
S_\pm(z)\equiv \frac{\sinh[(\hat{B}\pm B)\arctan z]}{\sinh[\pi(\hat{B}\pm B)/2]} \, , \\ 
T_\pm &\equiv& \coth\frac{\pi(\hat{B}+ B)}{2}\pm \coth\frac{\pi(\hat{B}- B)}{2} \, .
\eea
\end{subequations}
Note that $T_\pm= C_+(\infty)\pm C_-(\infty)=C_+(-\infty)\pm C_-(-\infty)$. Since the functions $C_\pm(z)$ and $S_\pm(z)$ are symmetric
and antisymmetric in $z$, respectively, both temporal components of the gauge fields are symmetric while both spatial components
are antisymmetric. Together with the behavior of the supercurrents under a parity transformation this ensures that the gauge fields transform
as a vector under parity, see discussion below Eq.\ (\ref{kF}). 
We plot the gauge fields with the supercurrents determined from minimization of the free energy, see Eqs.\ (\ref{jmin}),
in the left panel of Fig.\ \ref{figfields}.
 
\FIGURE[t]{\label{figfields}
{ \centerline{\def\epsfsize#1#2{0.7#1}
\epsfbox{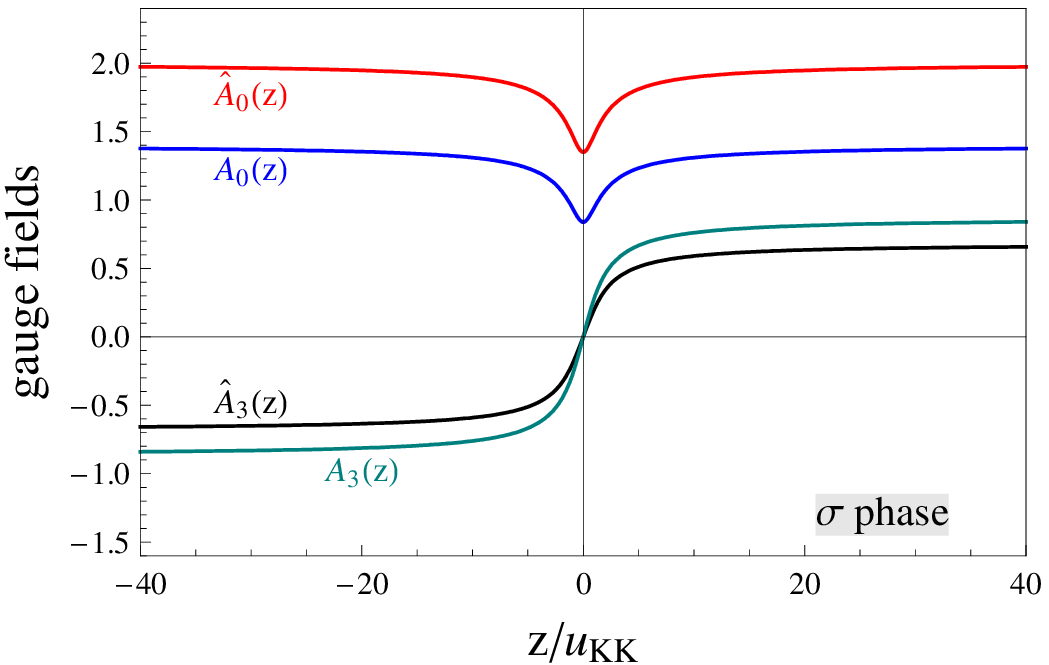}
\epsfbox{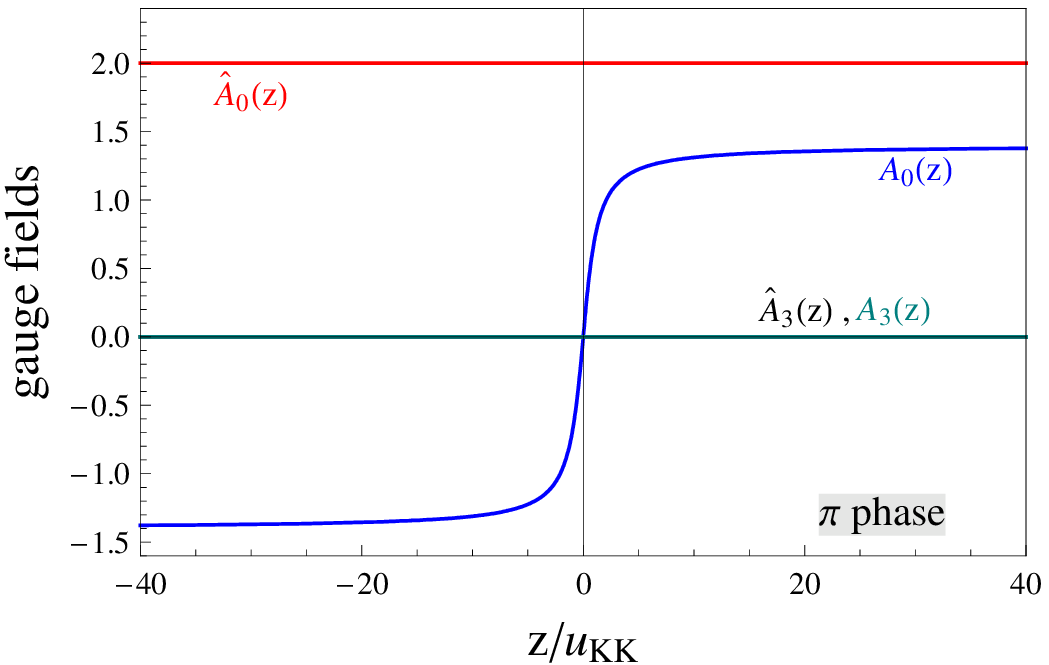}}
}
\caption{ 
Energetically preferred configuration of the gauge fields as a function of the holographic coordinate $z$ 
for the sigma phase (left panel) and the charged
pion phase (right panel). For the sigma phase we have chosen a dimensionless magnetic field $eB_{\rm em}=2$.
In the pion phase, $B_{\rm em}=0$ due to the Meissner effect.  
The boundary values for $\hat{A}_0(z)$ and $A_0(z)$ are given by (twice) the baryon and isospin
 chemical potentials, respectively. The boundary values of $\hat{A}_3(z)$ and $A_3(z)$ yield the meson
supercurrents and are determined dynamically from minimization of the free energy.
 }  
}

Next, we insert the gauge fields and the resulting field strengths into the free energy (\ref{Omegaconf}). We drop the contributions 
$\Omega_g$ and, via holographic renormalization, $\Omega_b$, as explained in Sec.\ \ref{sec:free}. Then we  
obtain (for details see Appendix \ref{appsigma})
\bea \label{Omegasigma2}
\Omega &=& \frac{2\kappa M_{\rm KK}^2}{3}\left[(\hat{\jmath}+\jmath)^2\rho_+(\hat{B},B)+(\hat{\jmath}-\jmath)^2\rho_-(\hat{B},B)\right.\non
&& \left. \hspace{1cm} -\, 4\mu_B(\hat{\jmath}\hat{B}+\jmath B)-4\mu_I(\hat{\jmath}B+\jmath\hat{B})\right] 
\, ,
\eea
with 
\be \label{rhopm}
\rho_\pm(\hat{B},B)\equiv 2(\hat{B}\pm B)\coth\frac{\pi(\hat{B}\pm B)}{2}+\frac{\pi(\hat{B}\pm B)^2}{2\sinh^2[\pi(\hat{B}\pm B)/2]} \, .
\ee
The asymptotic values of the functions $\rho_\pm(\hat{B},B)$ at small and large magnetic fields are shown in 
Table \ref{tablerhoeta} in Appendix \ref{appsigma}. 

Minimizing $\Omega$ with respect to $\hat{\jmath}$, $\jmath$ yields
\begin{subequations} \label{jmin}
\bea
\hat{\jmath}&=&\frac{\mu_B+\mu_I}{2}\frac{\hat{B}+B}{\rho_+(\hat{B},B)}+\frac{\mu_B-\mu_I}{2}\frac{\hat{B}-B}{\rho_-(\hat{B},B)} \, , \\
\jmath&=&\frac{\mu_B+\mu_I}{2}\frac{\hat{B}+B}{\rho_+(\hat{B},B)}-\frac{\mu_B-\mu_I}{2}\frac{\hat{B}-B}{\rho_-(\hat{B},B)} \, .
\eea
\end{subequations}
One can check that this is indeed a minimum of $\Omega$:
the matrix of second derivatives of $\Omega$ with respect to the supercurrents has eigenvalues 
$8\kappa M_{\rm KK}^2\rho_+/3$, $8\kappa M_{\rm KK}^2\rho_-/3$, which are independent of 
$\hat{\jmath}$ and $\jmath$ and positive for all $\hat{B}$, $B$. 

As already mentioned below Eq.\ (\ref{ansatzb}) we recall that the supercurrents $\hat{\jmath}$, $\jmath$ act as a source for the normal currents  
$\hat{\cal J}$, ${\cal J}$ which are the spatial 3-components of the four-currents
\be \label{defcalJ}
{\cal J}_{R/L}^\mu = {\cal J}_{{\rm YM},R/L}^{\mu} + {\cal J}_{{\rm CS},R/L}^{\mu} \, ,
\ee
where the Yang-Mills and Chern-Simons contributions are given by \cite{Hashimoto:2008zw,Hata:2008xc}
\begin{subequations} \label{defJ}
\bea 
{\cal J}_{{\rm YM},R/L}^{\mu} &=& \mp 2\kappa M_{\rm KK}^2k(z){\cal F}^{\mu z}\Big|_{z=\pm\infty} \, , \label{defJYM} \\
{\cal J}_{{\rm CS},R/L}^{\mu} &=& \mp \frac{N_c}{24 \pi^2}\epsilon^{\mu\nu\rho\sigma} {\cal A_\nu}{\cal F}_{\rho\sigma}\Big|_{z=\pm \infty} \, .
\eea
\end{subequations}
Here, the indices $\mu,\nu,\rho,\sigma$ run over 0,1,2,3, the upper (lower) signs correspond to $R$ ($L$), 
and we have, in the Chern-Simons contribution, already used that in our ansatz the off-diagonal 
components of the gauge fields in flavor space vanish. With the gauge fields (\ref{gaugefields1}) and the field strengths (\ref{kF})
we obtain the baryon and isospin components of the spatial currents,
\begin{subequations} \label{JJ}
\bea
\hat{\cal J}&\equiv& \hat{\cal J}_{R} = -\hat{\cal J}_{L}
= \kappa M_{\rm KK}^2\left[(\mu_B+\mu_I)\frac{(\hat{B}+B)^2}{\rho_+(\hat{B},B)}\coth\frac{\pi(\hat{B}+B)}{2} \right. \non
&&\left. 
+\,(\mu_B-\mu_I)\frac{(\hat{B}-B)^2}{\rho_-(\hat{B},B)}\coth\frac{\pi(\hat{B}-B)}{2}-\frac{2}{3}(\mu_B\hat{B}+\mu_I B)\right] \, , \qquad \\[2ex] 
\cal J&\equiv& {\cal J}_{R} = -{\cal J}_{L}= 
\kappa M_{\rm KK}^2\left[(\mu_B+\mu_I)\frac{(\hat{B}+B)^2}{\rho_+(\hat{B},B)}\coth\frac{\pi(\hat{B}+B)}{2} \right.\non
&&\left. -\,(\mu_B-\mu_I)\frac{(\hat{B}-B)^2}{\rho_-(\hat{B},B)}\coth\frac{\pi(\hat{B}-B)}{2}
-\frac{2}{3}(\mu_B B+\mu_I\hat{B})\right]\, , \qquad 
\eea
\end{subequations}
where the terms with prefactor 2/3 are the Chern-Simons contributions.
These currents are already evaluated at the minimum of the free energy, i.e., we have inserted the supercurrents (\ref{jmin}).
They add up to zero in the sums ${\mathcal J}_L+{\mathcal J}_R$
and $\hat{\mathcal J}_L+\hat{\mathcal J}_R$, 
corresponding to vanishing baryon and isospin currents,
however they yield
nonzero axial currents.

For small magnetic fields, the linear terms of the Yang-Mills
and Chern-Simons contributions cancel exactly and thus the currents become cubic in the magnetic field,
\begin{subequations} \label{currentsJ}
\bea
\hat{\cal J} &\simeq& \frac{\kappa M_{\rm KK}^2 \pi^2}{54}\left[(\hat{B}+B)^3(\mu_B+\mu_I)+(\hat{B}-B)^3(\mu_B-\mu_I)\right] \, , \\
{\cal J} &\simeq& \frac{\kappa M_{\rm KK}^2 \pi^2}{54}\left[(\hat{B}+B)^3(\mu_B+\mu_I)-(\hat{B}-B)^3(\mu_B-\mu_I)\right] \, .
\eea
\end{subequations}
The cancellation of the linear terms seems to suggest that there might also be cancellations in the cubic terms from terms we have neglected
upon expanding the DBI action for small gauge fields, possibly leading to vanishing currents. 
By considering a one-flavor system (where the use of the full DBI action is much simpler)
we have checked that this is not the case.\footnote{This is at variance with 
Ref.\ \cite{Bergman:2008qv} where a vanishing 
axial current has been found. In this reference, however, a modified action has been used where certain surface terms are dropped,
essentially to {\it force} the current to vanish.
Since we do not see any contradiction in a nonvanishing current (in the presence of an external magnetic field) we leave the 
full resolution of this discrepancy to future studies.}

We plot the currents $\hat{\cal J}$, ${\cal J}$ in Fig.\ \ref{figcurrents}, where we also show the supercurrents and the densities in the
sigma phase. We remark that the expansion of the 3-component of the gauge fields does not contain the complete current. One rather only finds the 
Yang-Mills contribution as a coefficient in front of the next-to-leading term
of the asymptotic expansion,
\bea \label{J1}
\hat{A}_3(z)  = \pm 2\hat{\jmath}\mp\frac{\hat{\cal J}_{{\rm YM},R/L}}{\kappa M_{\rm KK}^2}\,\frac{1}{z}+{\cal O}\left(\frac{1}{z^2}\right) \, , 
\quad\; 
A_3(z)  = \pm 2\jmath\mp\frac{{\cal J}_{{\rm YM},R/L}}{\kappa M_{\rm KK}^2}\,\frac{1}{z}+{\cal O}\left(\frac{1}{z^2}\right) \, . \quad
\eea

Finally, inserting the values (\ref{jmin}) back into $\Omega$ yields the value of the free energy at the minimum,
\bea \label{S02}
\Omega_\sigma &=& -\frac{2\kappa M_{\rm KK}^2}{3}
\left[(\mu_B+\mu_I)^2\frac{(\hat{B}+B)^2}{\rho_+(\hat{B},B)}+(\mu_B-\mu_I)^2\frac{(\hat{B}-B)^2}{\rho_-(\hat{B},B)}\right] \, .
\eea
We see that for vanishing magnetic fields $\Omega_\sigma=0$, i.e., the free energy does not depend on any of the chemical potentials.
This is the expected result for the sigma phase and has also been observed in Ref.\ \cite{Parnachev:2007bc}. 

%Note that in this 
%reference also phases with $\langle\bar{u} u\rangle$ different from $\langle\bar{d} d\rangle$ have been considered (more precisely, phases
%where one of these condensates vanishes). This possibility
%is indeed expected in a situation with nonzero isospin chemical potential \cite{Klein:2003fy}. 
%In Ref.\ \cite{Parnachev:2007bc} a phase with $\langle\bar{u} u\rangle \neq 0$, $\langle\bar{d} d\rangle = 0$ (or vice versa) is 
%realized within the deconfined but
%chirally broken phase, which does not exist in our present setup with maximally separated D8 and $\overline{\rm D8}$ branes.

%%%%%%%%%%%%%%%%%%%%%%%%%%%%%%%%%%%%%%%%%%%%%%%%%%%%%%%%%%%%%%%%%%%%%%%%%%%%%%%%%%%%%%%%%%%%
\subsubsection{Pion phase and Meissner effect}
\label{sec:meissner}
%%%%%%%%%%%%%%%%%%%%%%%%%%%%%%%%%%%%%%%%%%%%%%%%%%%%%%%%%%%%%%%%%%%%%%%%%%%%%%%%%%%%%%%%%%%%

In this case the boundary conditions are given in the second row of Table \ref{tableboundary}, and the differential equations (\ref{magn}) for
the magnetic fields have the solution 
\bea \label{twist}
\hat{b}(z) = \hat{\cal B} \, , \qquad b(z)=\frac{2{\cal B}}{\pi}\arctan z \, . 
\eea
As we have discussed at the end of Sec.~\ref{sec:free}, nonconstant
functions $\hat b$ or $b$ lead to an infinite contribution to the
free energy which cannot be removed by holographic renormalization, but
which enforces a vanishing magnetic field, indicating a Meissner effect.
In the $\pi$ phase, it is
the nonconstant isospin component $b(z)$ 
which leads to this conclusion. This is only to be expected
since 
%seems to indicate the presence of a curious term in the free energy which is proportional to the 
%square of the extension of the system in the direction perpendicular to the magnetic field, cf.\ Eq.\ (\ref{OmegaH}). Inserting 
%the solution (\ref{twist}) into $\Omega_b$ from Eq.\ (\ref{OmegaH}) shows that, in the second term, 
%the integral over the holographic coordinate $z$ remains finite, while the 
%integral over three-dimensional space yields an infinite (positive) energy density in the infinite volume limit.   
%This apparent problem is solved because we must expect a Meissner effect for the pion condensed phase. The 
the condensate of pions 
carries an electric charge, and thus the system is an electromagnetic superconductor. By the Meissner effect, a magnetic field is induced which is opposite, 
but equal in magnitude, to the applied magnetic
field, such that ${\cal B}_{\rm em}=0$ and thus $\hat{B}=B=0$. Of course our electromagnetic group is only global and thus 
the microscopic description of the Meissner effect, for instance in terms of a Meissner mass for the photon, is not straightforward.
However, in terms of supercurrents, the effect can be described quite naturally: in fact we have to allow for a supercurrent 
in the directions transverse to the magnetic field, i.e., $\bm{\jmath}_s({\bf x},z) = \frac12 b(z)\,(x_2,-x_1,0)$, such that 
${\rm curl}\,\bm{\jmath}_s=-b$ (and the same for the components $\hat{\bm\jmath}_s$, $\hat{b}$). 
This is the usual London equation for a superconductor, 
see for instance Ref.\ \cite{fetter}. 
Consequently, we need to add the supercurrents $\hat{\bm\jmath}_s$, $\bm{\jmath}_s$ to the boundary conditions of the gauge fields 
$\hat{A}_1({\bf x},z)$, $\hat{A}_2({\bf x},z)$, 
$A_1({\bf x},z)$, $A_2({\bf x},z)$ from Eqs.\ (\ref{ansatzb})
such that the total boundary conditions (and thereby the total magnetic field in the superconductor) vanish, $\hat{B}=B=0$. 
This condition renders the equations of motion for the pion phase very simple. We shall, however, solve
these equations for arbitrary magnetic fields and only at the end set $\hat{B}=B=0$. This provides us with a better understanding 
of the structure of the solution, for instance its behavior under parity transformations. 

We defer all technical details and the solution for general boundary conditions with the magnetic fields (\ref{twist}) to Appendix \ref{apppi}.
For the specific boundary conditions characterizing the charged pion condensate we find the solutions
\begin{subequations} \label{gaugefields2}
\bea
\hat{A}_0(z) &=& 2\mu_B +  \mu_I[\tilde{C}_+(z)+\tilde{C}_-(z)-\tilde{T}_+] + \hat{\jmath}[\tilde{C}_+(z)-\tilde{C}_-(z)-\tilde{T}_-]  \, , \\
A_0(z) &=& \mu_I[\tilde{S}_+(z)+\tilde{S}_-(z)] +\hat{\jmath}[\tilde{S}_+(z)-\tilde{S}_-(z)] \, , \\
\hat{A}_3(z) &=& \hat{\jmath}[\tilde{S}_+(z)+\tilde{S}_-(z)]+\mu_I[\tilde{S}_+(z)-\tilde{S}_-(z)]  \, , \\
A_3(z) &=&\hat{\jmath}[\tilde{C}_+(z)+\tilde{C}_-(z)-\tilde{T}_-]
+\mu_I[\tilde{C}_+(z)-\tilde{C}_-(z)-\tilde{T}_+] \, , 
\eea
\end{subequations}
where we abbreviated
\begin{subequations} \label{CSTtilde}
\bea
\tilde{C}_+(z)&\equiv& \frac{P_+(z)+P_-(z)}{P_+^+-P_+^-} \, , \qquad \tilde{C}_-(z)\equiv \frac{Q_+(z)+Q_-(z)}{Q_+^+-Q_+^-} \, , \\
\tilde{S}_+(z)&\equiv& \frac{P_+(z)-P_-(z)}{P_+^+-P_+^-} \, , \qquad \tilde{S}_-(z)\equiv \frac{Q_+(z)-Q_-(z)}{Q_+^+-Q_+^-} \, , \\
\tilde{T}_\pm&\equiv& \frac{P_+^++P_+^-}{P_+^+-P_+^-} \pm \frac{Q_+^++Q_+^-}{Q_+^+-Q_+^-} \, ,
\eea
\end{subequations}
with 
\begin{subequations} \label{PQ}
\bea
Q_\pm(z)&\equiv& \frac{\pi}{2\sqrt{B}}e^{\frac{\pi\hat{B}^2}{4B}}{\rm erf}\left(\frac{\pi\hat{B}\pm 2B\arctan z}{2\sqrt{\pi B}}\right) \, , \\
P_\pm(z)&\equiv& \frac{\pi}{2\sqrt{B}}e^{-\frac{\pi\hat{B}^2}{4B}}{\rm erfi}\left(\frac{\pi\hat{B}\pm 2B\arctan z}{2\sqrt{\pi B}}\right) \, , 
\eea
\end{subequations}
and $Q_+^\pm\equiv Q_+(\pm\infty)$, $P_+^\pm\equiv P_+(\pm\infty)$. Here, erf is the error function and ${\rm erfi}(z)\equiv {\rm erf}(iz)/i$.
The functions $\tilde{C}_\pm$, $\tilde{S}_\pm$, $\tilde{T}_\pm$
are the more complicated counterparts of the functions $C_\pm$, $S_\pm$, $T_\pm$ from Eqs.\ (\ref{CST}). They share the same property 
$\tilde{T}_\pm= \tilde{C}_+(\infty)\pm \tilde{C}_-(\infty)=\tilde{C}_+(-\infty)\pm \tilde{C}_-(-\infty)$, and, as their 
counterparts, $\tilde{C}_\pm(z)$ and $\tilde{S}_\pm(z)$ are symmetric and antisymmetric in $z$, respectively. This means that the 
temporal ${\bf 1}$-component and the spatial $\tau_3$-component are symmetric in $z$ while the temporal $\tau_3$-component 
and the spatial-${\bf 1}$ component are antisymmetric. Again, together with the parity transformations of the supercurrents, this gives the 
correct parity behavior of the gauge fields, see discussion in Appendix \ref{apppi}. 
In particular, the requirement of a well-defined parity leads to the condition $\jmath=0$.
%We plot the gauge fields with the currents at the energetic minimum, see Eq.\ (\ref{jmin1}),
%in the right panel of Fig.\ \ref{figfields}.

Inserting Eqs.\ (\ref{gaugefields2}) and the corresponding field strengths into the free energy (\ref{Omegaconf}) yields
\bea \label{Omegapion}
\Omega &=& \frac{2\kappa M_{\rm KK}^2}{3}\left\{(\hat{\jmath}^2-\mu_I^2)\,\rho(\hat{B},B) 
-2\mu_B[\mu_I\,\eta_+(\hat{B},B)+\hat{\jmath}\,\eta_-(\hat{B},B)]\right\} 
+\Omega_b
\, ,\quad  
\eea
with $\Omega_b$ given in Eq.\ (\ref{OmegaH}) and
\begin{subequations} \label{rhoeta}
\bea
\rho(\hat{B},B)&\equiv& \frac{4\pi}{(P_+^+-P_+^-)(Q_+^+-Q_+^-)} +4\cosh\frac{\pi\hat{B}}{2}\left(\frac{e^{\frac{\pi B}{4}}}{P_+^+-P_+^-}+
\frac{e^{-\frac{\pi B}{4}}}{Q_+^+-Q_+^-}\right) \, , \\
\eta_\pm(\hat{B},B)&\equiv&2\sinh\frac{\pi\hat{B}}{2}\left(\frac{e^{\frac{\pi B}{4}}}{P_+^+-P_+^-}\mp
\frac{e^{-\frac{\pi B}{4}}}{Q_+^+-Q_+^-}\right) \, .
\eea
\end{subequations}
Again, the asymptotic values of these functions are given in Table \ref{tablerhoeta} in Appendix \ref{appsigma}.
 
Minimization of $\Omega$ with respect to $\hat{\jmath}$ yields
\be \label{jmin1}
\hat{\jmath}=\mu_B\frac{\eta_-(\hat{B},B)}{\rho(\hat{B},B)} \, ,  
\ee
and the minimum of the free energy becomes
\be \label{Omega0pi}
\Omega_\pi = -\frac{2\kappa M_{\rm KK}^2}{3} \left[\mu_B^2\frac{\eta_-^2(\hat{B},B)}{\rho(\hat{B},B)}+\mu_I^2\,\rho(\hat{B},B)
+2\mu_B\mu_I\,\eta_+(\hat{B},B)
\right]  +\Omega_b
\, .
\ee
We see that for vanishing magnetic fields the free energy depends on the isospin chemical potential, giving rise to a 
nonzero isospin density. This is expected from the quark content of the charged pion condensate and was also observed within the 
Sakai-Sugimoto model in Ref.\ \cite{Parnachev:2007bc}. 

Taking into account the Meissner effect, which is enforced by
the infrared divergence in $\Omega_b$ 
as discussed at the beginning of this subsection, we have to set
$\hat{B}=B=0$, leading to the simple result for the free energy
\be \label{Ompi}
\Omega_\pi = -\frac{8\kappa M_{\rm KK}^2}{\pi}\, \mu_I^2  \, .
\ee
From Eq.\ (\ref{jmin1}) we conclude that for $\hat{B}=B=0$ we have $\hat{\jmath}=0$. 
And, from the definitions (\ref{defJ}), we see that in the absence of a magnetic
field also the normal currents vanish, $\hat{\cal J}={\cal J}=0$. In the following we shall discuss the results of the 
pion phase only in the presence of the Meissner effect.

%%%%%%%%%%%%%%%%%%%%%%%%%%%%%%%%%%%%%%%%%%%%%%%%%%%%%%%%%%%%%%%%%%%%%%%%%%%%%%%%%%%%%%%%%%%%
\subsection{Meson supercurrents and number densities} 
\label{sec:properties1}
%%%%%%%%%%%%%%%%%%%%%%%%%%%%%%%%%%%%%%%%%%%%%%%%%%%%%%%%%%%%%%%%%%%%%%%%%%%%%%%%%%%%%%%%%%%%

We have seen that all currents in the charged
pion phase vanish (except for the supercurrents in the transverse 1- and 2-directions which 
cancel the applied magnetic field). The supercurrents and normal currents in the sigma phase, given in Eqs.\ (\ref{jmin}) and (\ref{JJ}), 
respectively, are shown in the left panel of Fig.\ \ref{figcurrents} as a function of the magnetic field. We have used the electromagnetic
field $B_{\rm em}$, defined in Eq.\ (\ref{Bem}), with the electric charges of up and down quarks. We see that the 
supercurrents behave linear 
in $B_{\rm em}$ for small $B_{\rm em}$ and approach an asymptotic value for a large magnetic field. These limit cases assume 
very simple forms in terms of the electric quark charges $q_i$ and the quark chemical potentials 
$\mu_{1,2}\equiv \mu_B\pm\mu_I$. The quark supercurrents $\jmath_{1,2}\equiv \hat{\jmath}\pm \jmath$ in the sigma phase then are 
\be
\jmath_i^\sigma \simeq
\frac{1}{2}\left\{ \begin{array}{cc} \displaystyle{\frac{\pi q_i\mu_iB_{\rm em}}{3}} & \;\;\mbox{for small}\; B_{\rm em} \\[2ex]
\displaystyle{\mu_i\,{\rm sgn}\,q_i}
& \;\;\mbox{for large}\; B_{\rm em} \end{array}\right. \, .
\ee
The limit of a large magnetic field is 
strictly speaking not consistent with our approximations. Firstly, we have expanded the DBI action for small gauge fields. Secondly, 
we have treated the flavor branes as probe branes which becomes questionable for large magnetic fields since one would have to consider the 
backreaction on the background geometry.
As mentioned in Ref.\ \cite{Thompson:2008qw} in the same context, the case of a large magnetic field within the present approach 
can only be meaningful if one thinks of the action (\ref{totalS}) as a ``bottom-up'' model for QCD, which is not derived from 
an AdS/CFT correspondence. 
Indeed, with appropriate functions $k(z)$, $h(z)$, the bottom-up model from Ref.\ \cite{Son:2003et} can be recovered 
from Eq.\ (\ref{YM5}). Thus, in the following we shall use our analytical functions to discuss the whole range of magnetic fields with this
qualification in mind.

\FIGURE[t]{\label{figcurrents}
{ \centerline{\def\epsfsize#1#2{0.7#1}
\epsfbox{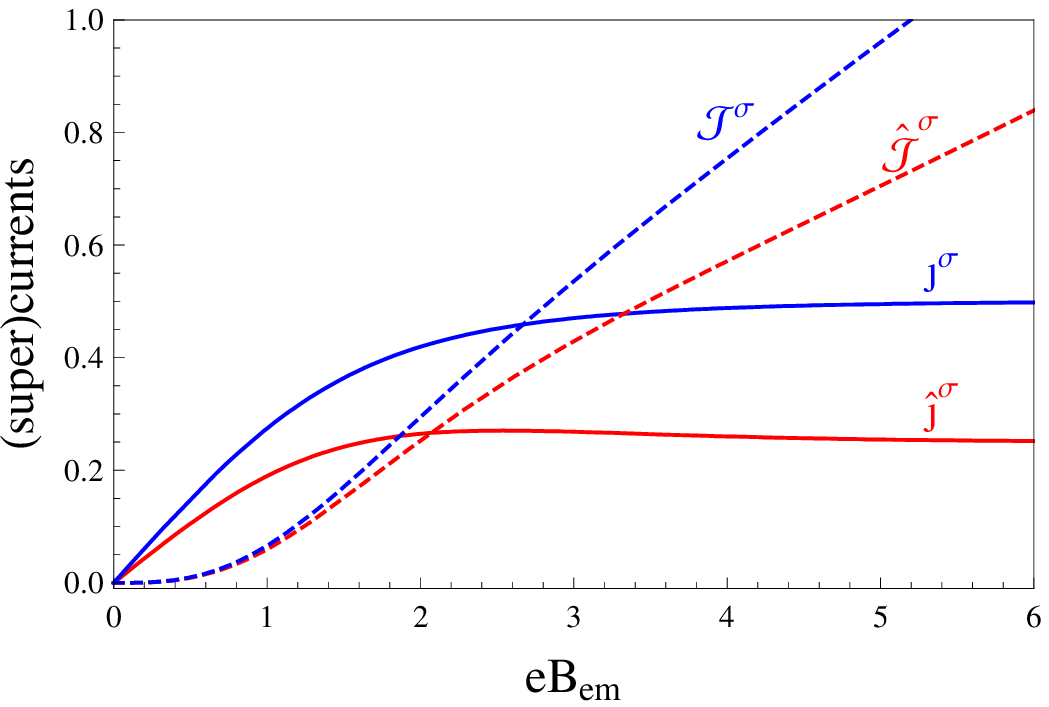}
\epsfbox{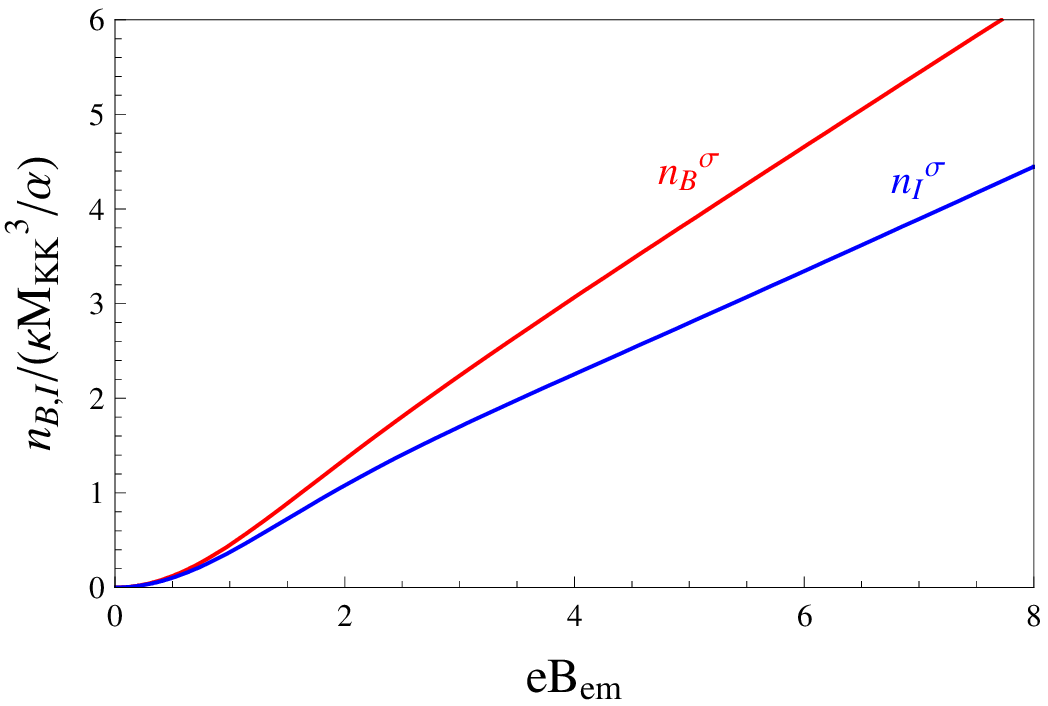}}
}
\caption{%(Color online) 
Left panel: Meson supercurrents $\hat{\jmath}^\sigma$ (red solid) and $\jmath^\sigma$ (blue solid) and normal currents 
$\hat{\cal J}^\sigma$ (red dashed) and ${\cal J}^\sigma$ (blue dashed) as a function of the dimensionless magnetic field 
$eB_{\rm em}$ in the sigma phase.  The units are 
$M_{\rm KK}/\alpha$ for $\hat{\jmath}^\sigma$, $\jmath^\sigma$ and $\kappa M_{\rm KK}^3/\alpha$ for $\hat{\cal J}^\sigma$, ${\cal J}^\sigma$. 
Right panel: baryon and isospin number densities as a function of the 
dimensionless magnetic field in the sigma phase. 
The analytical expressions for the functions are given in Eqs.\ (\ref{jmin}), (\ref{JJ}), and (\ref{denssigma}). 
We have fixed $\mu_B=2\mu_I=M_{\rm KK}\alpha$. 
In the (charged) pion phase, all currents 
as well as the baryon density vanish due to the Meissner effect; the isospin density is given by the simple expression (\ref{denspiI}).    }
}

We can compute the baryon and isospin densities from the free energies computed in the previous sections via 
\be \label{defn}
n_{B,I}=-\frac{\partial\Omega}{\partial \mu_{B,I}} \, .
\ee
We obtain for the sigma phase
\begin{subequations} \label{denssigma}
\bea
n_B^\sigma &=&  \frac{4\kappa M_{\rm KK}^2}{3}\left[(\mu_B+\mu_I)\frac{(\hat{B}+B)^2}{\rho_+}+
(\mu_B-\mu_I)\frac{(\hat{B}-B)^2}{\rho_-}\right] \, , \label{denssigma1} \\
n_I^\sigma &=&  \frac{4\kappa M_{\rm KK}^2}{3}\left[(\mu_B+\mu_I)\frac{(\hat{B}+B)^2}{\rho_+}-
(\mu_B-\mu_I)\frac{(\hat{B}-B)^2}{\rho_-}\right] \, ,
\eea
\end{subequations}
%The densities can as well be obtained from the 0-component of the four-current defined in Eq.\ (\ref{defcalJ}), i.e., one can check that
which agrees with the 0-component of the four-current defined in Eq.\ (\ref{defcalJ}),
\be\label{J0dOdmu}
{\cal J}_{R}^0 + {\cal J}_{L}^0 = -\frac{\partial\Omega}{\partial \mu_{B}} -\frac{\partial\Omega}{\partial \mu_{I}} \, \tau_3 \, ,
\ee
provided we use $2\mu_{B,I}$ (as opposed to $\mu_{B,I}$) for the boundary values of $\hat{A}_0$, $A_0$, see Table \ref{tableboundary}.
Remarkably,
with these simple rescalings we
can reconcile the  thermodynamic interpretation of $\Omega$ as a free energy
with the standard definition of the currents from gauge/gravity duality.
%confirming the thermodynamic interpretation of $\Omega$ as a free energy. 
The reason for this unconventional normalization appears to be the anomalous nature of baryon and isospin number in the present model\footnote{This
issue is in fact related to the remark below Eq.\ (\ref{currentsJ}): 
using a modified action as in Ref.\ \cite{Bergman:2008qv} which
drops certain surface terms would allow us to use the more natural boundary values $\mu_{B,I}$.  
We have checked that, apart from the difference in the axial current, 
the use of the modified action does not lead to qualitative changes in our following results. 
More precisely, changing the action by hand as in Ref.\ \cite{Bergman:2008qv} amounts to multiplying the Chern-Simons action
by 3/2, and rescaling the chemical potential by a factor 1/2. For the phase diagram presented in Sec.\ \ref{sec:compare}, only the latter
modification is essential and leads to corresponding quantitative changes.}.

As for the 3-components of the gauge fields (\ref{J1}), also the next-to-leading terms in the
asymptotic expansions of the 0-components
only contain the Yang-Mills contribution
$\mathcal J^0_{\rm YM}$ of Eq.~(\ref{defJYM}).

The densities are plotted in the right panel of Fig.\ \ref{figcurrents}. As expected, both densities
vanish in the case of a vanishing magnetic field. Switching on a magnetic field induces currents as well as nonzero
densities. Again, it is convenient to express the number densities in terms of the quark flavor components, $n_{1,2}\equiv n_B\pm n_I$, 
rather than in 
baryon and isospin components. We obtain for small and large magnetic fields
\be
n_i^\sigma \simeq
\frac{4\kappa M_{\rm KK}^2}{3}\left\{ \begin{array}{cc} \displaystyle{\frac{\pi q_i^2\mu_iB_{\rm em}^2}{3}} & \;\;\mbox{for small}\; B_{\rm em} 
\\[2ex]
\displaystyle{\mu_i|q_i|B_{\rm em}}
& \;\;\mbox{for large}\; B_{\rm em} \end{array}\right. \, .
\ee
For the pion phase we find from the free energy (\ref{Ompi})
\begin{subequations} \label{denspi}
\bea
n_B^\pi &=& 0 \, , \\
n_I^\pi &=& \frac{16\kappa M_{\rm KK}^2}{\pi}\mu_I \, . \label{denspiI}
\eea
\end{subequations}
From the baryon and isospin densities we can immediately deduce the electric charge density 
$n_Q=q_1n_1+q_2n_2$. 
The electric charge of the system is relevant for example in the astrophysical context because in a neutron star the overall 
electric charge has to vanish. Here we  
simply observe which electric charge is carried by our system for given chemical potentials. For more realistic applications one would have to 
require charge neutrality and possibly counterbalance the charge of the chiral condensate for instance by the presence of electrons
or protons. 
For the $\sigma$ phase we find $n_Q^\sigma=0$ for vanishing magnetic fields, as expected. Switching on a magnetic field induces electric
charges in the system. For infinitesimally small $B_{\rm em}$ a straight line $\mu_B=-9\mu_I/7$ appears in the $\mu_B$-$\mu_I$ plane 
dividing the plane into a region with infinitesimally positive (above/right of the line) and 
negative (below/left of the line) charge. With increasing magnetic field, giving rise to larger charges, the slope of the 
line slightly decreases and approaches the value $\mu_B=-5\mu_I/3$ asymptotically for large $B_{\rm em}$. 
For the pion phase we have $n_Q^\pi=n_I^\pi$, which is positive (negative) for positive (negative) isospin chemical potentials and 
independent of the baryon chemical potential.

We may finally recover the scenario considered in Ref.\ \cite{Thompson:2008qw} as a limit of our more general results. In that paper, 
a vanishing isospin chemical potential, a vanishing baryon component of the magnetic field, and an isospin magnetic field constant in the holographic
coordinate $z$ (as in Eq.\ (\ref{twist})) was considered. For a comparison it is thus instructive to compute the energy density 
$\epsilon$ of the sigma phase. We write the free energy (\ref{Omegasigma2}) as $\Omega = \epsilon -\mu_B n_B- \mu_I n_I $
with $n_B$, $n_I$ given in Eqs.\ (\ref{denssigma}). Then we can express the energy density in terms of the number densities,
\be
\epsilon_\sigma= 
\frac{3}{32\kappa M_{\rm KK}^2}\left[(n_B^\sigma+n_I^\sigma)^2\frac{\rho_+}{(\hat{B}+B)^2}+
(n_B^\sigma-n_I^\sigma)^2\frac{\rho_-}{(\hat{B}-B)^2}\right] \, .
\ee
For small and large magnetic fields we obtain
\be \label{energydens}
\epsilon_\sigma \simeq
\left\{ \begin{array}{cc} \displaystyle{\frac{2\lambda M_{\rm KK}^2}{3N_c}
\left[\frac{(n_B^\sigma+n_I^\sigma)^2}{(\hat{\cal B}+{\cal B})^2}+\frac{(n_B^\sigma-n_I^\sigma)^2}{(\hat{\cal B}-{\cal B})^2}
\right] }
 & \;\;\mbox{for small}\; \hat{\cal B},{\cal B} \\[4ex]
\displaystyle{\frac{3\pi^2}{N_c}\left[\frac{(n_B^\sigma+n_I^\sigma)^2}{|\hat{\cal B}+{\cal B}|}
+\frac{(n_B^\sigma-n_I^\sigma)^2}{|\hat{\cal B}-{\cal B}|}
\right]  }
& \;\;\mbox{for large}\; \hat{\cal B},{\cal B}\end{array}\right. \, ,
\ee
where we have reinstated the dimensionful magnetic fields according
to Eq.\ (\ref{dimless}). For large magnetic fields we thus obtain, up to a numerical prefactor, an equation of state   
as for a free fermion gas in a magnetic field: setting $n_I^\sigma=\hat{\cal B}=0$ we have $\epsilon_\sigma = 6\pi^2(n_B^\sigma)^2/({\cal B}N_c)$
while for a free gas $\epsilon_0=\pi^2 n_B^2/({\cal B}N_c)$ \cite{Thompson:2008qw}. 
In Ref.\ \cite{Thompson:2008qw} even the prefactor is exactly that of the free gas. 
We have checked that this discrepancy comes from the surface term in the Chern-Simons action:
had we dropped the contribution of the surface term (\ref{surface}) for the free energy, we would have reproduced Eqs.\ (34) and (35) 
of Ref.\ \cite{Thompson:2008qw} exactly (upon setting $n_I^\sigma=\hat{\cal B}=0$ in our result). 
We can also explain this discrepancy in another way. We have computed the free energy, taking into account the Yang-Mills and Chern-Simons
contributions, and then computed the baryon density (and isospin density) by the thermodynamic relation (\ref{defn}). Given this definition 
of $n_B$ we can write
\bea
n_B=\frac{N_c}{6\pi^2}(\hat{\jmath}\hat{\cal B}+\jmath {\cal B}) 
=\frac{N_c}{96\pi^2}\int_{-\infty}^\infty dz\,\epsilon_{MNPQ} {\rm Tr}[{\cal F}_{MN}{\cal F}_{PQ}] \, , 
\eea
where $M,N,P,Q = z,1,2,3$, where the first equality can be read off from Eq.\ (\ref{Omegasigma2}) 
upon reinstating the dimensionful magnetic field, and the second equality follows from the field strengths in Eqs.\ (\ref{kF}). 
The resulting expression on the right-hand side differs by a factor 1/3 from the one in 
Ref.\ \cite{Thompson:2008qw}; the latter is normalized such that the baryon number is equal to the instanton number for a static 
instanton configuration \cite{Hata:2007mb} (in the absence of a magnetic field). 
Consequently, had we {\it required} this normalization we would have had to adjust the baryon chemical potential 
to be $3\mu_B$ with $\mu_B$ the baryon chemical potential used above. In this case we would have reproduced  the small $B$
limit for the energy density of Ref.\ \cite{Thompson:2008qw} exactly. However, in the large $B$ limit we would then have obtained
$\epsilon_\sigma = 2\epsilon_0/3$.  The reason is that, 
due to the surface term, the functional dependence of the energy density
on the magnetic field differs from Ref.\ \cite{Thompson:2008qw}, deviating by different factors in the two limits of small and large magnetic
fields. 

%%%%%%%%%%%%%%%%%%%%%%%%%%%%%%%%%%%%%%%%%%%%%%%%%%%%%%%%%%%%%%%%%%%%%%%%%%%%%%%%%%%%%%%%%%%%
\subsection{Phase diagram and critical magnetic field}
\label{sec:compare}
%%%%%%%%%%%%%%%%%%%%%%%%%%%%%%%%%%%%%%%%%%%%%%%%%%%%%%%%%%%%%%%%%%%%%%%%%%%%%%%%%%%%%%%%%%%%

In this section we determine which of the two phases is favored for given values of the chemical potentials and the magnetic 
field. To this end, notice that the holographic description of our system is our ``microscopic'' theory; therefore, 
we have identified the boundary values of the bulk field strengths as ${\cal B}_{\rm em}$, not as ${\cal H}_{\rm em}$. We are interested, however, 
in a free energy comparison at a fixed external magnetic field ${\cal H}_{\rm em}$. Consequently, we have to apply a Legendre transformation
to construct the Gibbs free energy $G$. (See for instance Ref.\ \cite{Noronha:2007wg} for an analogous construction.) 
In the case of the charged pion condensate, where ${\cal B}_{\rm em}=0$, the Gibbs free energy is identical 
to the above computed free energy,
\be \label{gibbspi}
G_\pi = \Omega_\pi = -\frac{8\kappa M_{\rm KK}^2}{\pi}\, \mu_I^2  \, .
\ee
It is convenient to introduce dimensionless free energies $\omega_{\sigma,\pi}$ via
\be
\Omega_{\sigma,\pi}=\frac{M_{\rm KK}^4}{\alpha^2}\,\kappa\,\omega_{\sigma,\pi}  \, . 
\ee
As we shall see below, $\kappa$ is the only parameter of the model on which the structure of the phase diagram depends, see Eq.\ (\ref{DG}). 
The other constants $M_{\rm KK}$ and $\alpha$ only set the energy scale. To make this $\kappa$ dependence explicit, 
we have pulled the dimensionless constant $\kappa$ out of $\omega_{\sigma,\pi}$. The dimensionless free energies are, 
using Eqs.\ (\ref{S02}) and (\ref{Ompi}),
\begin{subequations}
\bea
\omega_\pi&=& -\frac{8\tilde{\mu}_I^2}{\pi} \,  , \\
\omega_\sigma&=& -\frac{2}{3}
\left[(\tilde{\mu}_B+\tilde{\mu}_I)^2\frac{(\hat{B}+B)^2}{\rho_+(\hat{B},B)}+(\tilde{\mu}_B-\tilde{\mu}_I)^2
\frac{(\hat{B}-B)^2}{\rho_-(\hat{B},B)}\right] \, ,
\eea
\end{subequations}
where we have introduced the dimensionless chemical potentials
\be
\tilde{\mu}_B\equiv \frac{\alpha\mu_B}{M_{\rm KK}} \, , \qquad \tilde{\mu}_I\equiv \frac{\alpha\mu_I}{M_{\rm KK}} \, .
\ee
To obtain the Gibbs free energy in the sigma phase we add the contribution ${\cal B}_{\rm em}^2/2$ and
Legendre transform the free energy with respect to the change of variable ${\cal B}_{\rm em}\to {\cal H}_{\rm em}$. Consequently, 
\be \label{gibbssigma}
G_\sigma = \frac{1}{2}{\cal B}_{\rm em}^2 + \Omega_\sigma - {\cal B}_{\rm em}{\cal H}_{\rm em}\, .
\ee
For a given external field ${\cal H}_{\rm em}$ one determines ${\cal B}_{\rm em}$ from the
stationarity condition
\be \label{stationary}
0 = \frac{\partial G_\sigma}{\partial {\cal B}_{\rm em}}= {\cal B}_{\rm em} - {\cal M}_\sigma  - {\cal H}_{\rm em}\, ,
\ee
where we defined the magnetization in the sigma phase
\be
{\cal M}_\sigma = -\frac{\partial \Omega_\sigma}{\partial {\cal B}_{\rm em}} = \frac{M_{\rm KK}^2}{\alpha}\,\kappa\,M_\sigma \, . 
\ee
Here, the dimensionless magnetization is given by 
\be \label{Msigma}
M_\sigma = \frac{4}{3}\left[ q_1\tilde{\mu}_1^2\frac{\hat{B}+B}{\rho_+}
\left(1-\frac{\hat{B}+B}{2\rho_+}\frac{\partial\rho_+}{\partial\hat{B}}\right)+q_2\tilde{\mu}_2^2\frac{\hat{B}-B}{\rho_-}
\left(1-\frac{\hat{B}-B}{2\rho_-}\frac{\partial\rho_-}{\partial\hat{B}}\right)\right] \, ,
\ee
where we have used Eq.\ (\ref{Bem}) for the electromagnetic field, and expressed the derivatives with respect to $B$ through derivatives with 
respect to $\hat{B}$. 
Before coming back to the Gibbs free energy let us discuss the magnetization and the resulting magnetic susceptibility $\chi_\sigma$. 
To obtain $M_\sigma$ as a function of the external magnetic field we first solve Eq.\ (\ref{stationary}), which, in dimensionless quantities reads
\be \label{HBM}
H_{\rm em} = B_{\rm em} -\kappa\,M_\sigma \, ,
\ee
numerically for $B_{\rm em}$. 
(Here, the dimensionless field $H_{\rm em}$ is defined analogously to the field $B_{\rm em}$, see Eq.\ (\ref{dimless}).)
Then, we insert the solution back into Eq.\ (\ref{Msigma}). The result depends on $\kappa$, for which we have to choose a numerical value. In order
to get some numerical estimates from our following results we also need to assign values to the other parameters of the model. 
Following \cite{Sakai:2004cn,Sakai:2005yt,Hashimoto:2008zw} we shall use 
\be\label{fitparams}
\kappa=0.00745\, , \qquad M_{\rm KK} = 949\,{\rm MeV}\, ,
\ee 
which has been obtained from fits to the rho meson mass and the pion decay constant. 
From this value for $\kappa$ we obtain, with $N_c=3$ and Eq.\ (\ref{kappa}), 
the 't Hooft coupling $\lambda\approx 16.6$, and then, with Eq.\ (\ref{alpha}), 
$\alpha\approx 2.55$. 

The full numerical result for the magnetization is shown in Fig.\ \ref{figmagnetization}. 
Our result is in qualitative agreement with Ref.\ \cite{Bergman:2008qv}, where the magnetization 
was computed for a one-flavor system. (Note, however, that in this reference the boundary value of the field strength was interpreted as 
$H$, not $B$.)  
We see that the magnetization behaves linearly for small magnetic fields. The slope is the magnetic susceptibility, i.e., $M_\sigma \simeq 
\chi_\sigma H_{\rm em}$. Upon expanding Eq.\ (\ref{Msigma}) for small magnetic fields we find
\be \label{chi}
\chi_\sigma = \frac{2\pi}{9} \frac{q_1^2\tilde{\mu}_1^2+q_2^2\tilde{\mu}_2^2}{1-\frac{2\kappa\pi}{9}(q_1^2\tilde{\mu}_1^2+q_2^2\tilde{\mu}_2^2)}
\, .
\ee
Since we neither expect the susceptibility to diverge nor to change sign, this result can only be trusted for sufficiently
small chemical potentials, roughly speaking $\tilde{\mu}_i^2\ll 1/(\kappa e^2)$. 
Given the numerical value $(\kappa e^2)^{-1/2}\simeq 38$ and given that one unit of the quark chemical potential $\tilde{\mu_i}=1$ corresponds 
to $\mu_i\simeq 400\,{\rm MeV}$, this is not a severe restriction for realistic values of $\mu_i$. However,
this result shows that in principle one has to be careful with large chemical potentials in the present approximation where we 
not only have expanded the DBI action for 
small gauge fields but also neglected the backreaction of the branes to the background geometry. 

For large magnetic fields the magnetization saturates. From Eq.\ (\ref{Msigma}) we find the constant value
\be \label{MlargeB}
\lim_{H_{\rm em}\to \infty} M_\sigma = \frac{q_1\tilde{\mu}_1^2-q_2\tilde{\mu}_2^2}{3} \, , 
\ee
where we have used that $B>\hat{B}$ for a two-flavor system with up and down quarks.

\FIGURE[t]{\label{figmagnetization}
{ \centerline{\def\epsfsize#1#2{0.7#1}
\epsfbox{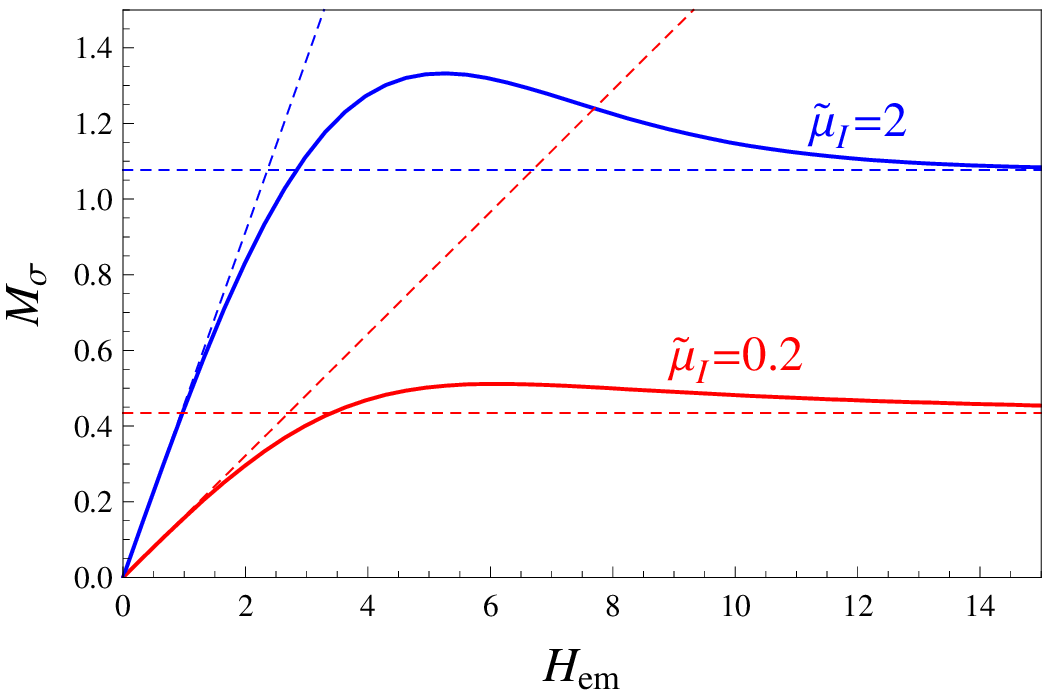}}
}
\caption{%(Color online) 
Dimensionless magnetization $M_\sigma$ for the sigma phase as a function of the dimensionless magnetic field $H_{\rm em}$ for 
two different values of the isospin chemical potential and a baryon chemical potential $\tilde{\mu}_B=2$. The dashed lines
are the susceptibilities $\chi_\sigma$ from Eq.\ (\ref{chi}), which approximate the magnetization for small magnetic fields, 
$M_\sigma = \chi_\sigma H_{\rm em}$, and the asymptotic values from Eq.\ (\ref{MlargeB}).}
}

We now return to the free energy comparison. For the sigma phase we insert Eq.\ (\ref{stationary}) into the Gibbs free energy 
(\ref{gibbssigma}). Then we obtain the following difference in Gibbs free energies
\be \label{DG}
\frac{\Delta G}{M_{\rm KK}^4/\alpha^2} \equiv \frac{G_\sigma-G_\pi}{M_{\rm KK}^4/\alpha^2}= -\frac{1}{2} H_{\rm em}^2+ \frac{1}{2}
\kappa^2 M_\sigma^2
+\kappa(\omega_\sigma - \omega_\pi) \, .
\ee
If $\Delta G >0$ ($\Delta G<0$) the $\pi$ ($\sigma$) phase is preferred.
It is instructive to relate the comparison between the sigma and pion phases to 
a usual superconducting material, say a metal,  where we compare the superconducting phase (corresponding to the pion phase) and the 
normal-conducting phase (corresponding to the sigma phase). With the help of this analogy we can understand the various terms in 
$\Delta G$. The term quadratic in $H_{\rm em}$ is negative, i.e., it works in favor of the normal-conducting phase. 
This term is the free energy cost which the superconducting phase has to pay
for creating a counter magnetic field in order to expel the external magnetic field. In a usual superconductor, this term is thus
responsible for a critical magnetic field beyond which the Cooper pair condensate breaks down. 
There is an additional term, working against the normal-conducting phase, 
proportional to $M_\sigma^2$. This 
term is absent in most usual superconductors which, to a good approximation, have no magnetic properties in their normal-conducting phase. 
We thus expect a competition between the two
terms, i.e., between the costs that the sigma and charged pion phase have to pay for the magnetization and the Meissner effect, respectively. 
This competition, together with the difference $\omega_\sigma - \omega_\pi$, will determine the resulting phase diagram. 

For small magnetic fields, $H_{\rm em}\ll 1$, and dimensionless chemical potentials of order one, $\tilde{\mu}_{B,I}\lesssim 1$,
 we can discuss the phase transition between the sigma and the charged pion phase 
analytically. In this case, the term $\kappa^2 M_\sigma^2$ is of the order of $\kappa^2 e^4$ and thus negligible. The free energy of
the sigma phase $\kappa\, \omega_\sigma$ is proportional to $\kappa \,e^2$ and thus also small compared to the remaining terms. 
We are left with the simple result
\be \label{DGapprox}
\frac{\Delta G}{M_{\rm KK}^4/\alpha^2} \simeq -\frac{1}{2} H_{\rm em}^2+ \frac{8\kappa}{\pi}\tilde{\mu}_I^2 \, .
\ee
At the phase transition $\Delta G=0$ we thus have  
\be \label{lines}
\tilde{\mu}_I = \pm \sqrt{\frac{\pi}{\kappa}}\frac{H_{\rm em}}{4} \approx \pm 5.13 \, H_{\rm em}\, .
\ee
This relation can be read as an equation for the critical magnetic field for a given isospin chemical potential, or as an 
equation for the critical chemical potential for a given magnetic field. We see that, in this approximation, the phase transition
is independent of the baryon chemical potential. In Fig.\ \ref{figHcrit} we plot the critical magnetic field as a function
of $\mu_I$ for two different values of $\mu_B$. 
%We see that the approximation from Eq.\ (\ref{lines})
%works quite well also for large magnetic fields.

\FIGURE[t]{\label{figHcrit}
{ \centerline{\def\epsfsize#1#2{0.7#1}
\includegraphics[width=0.5\textwidth]{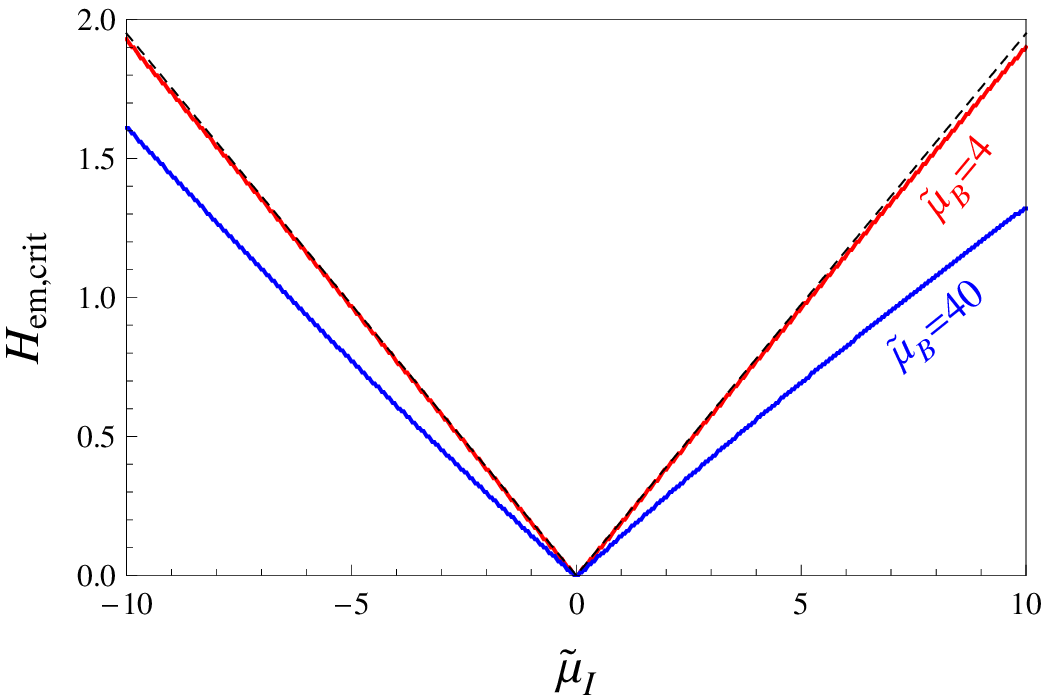}}
}
\caption{Critical magnetic field for the phase transition from the pion to the sigma phase as a function of the isospin
chemical potential for baryon chemical potentials $\tilde{\mu}_B=4$ (red curve) and $\tilde{\mu}_B=40$ (blue curve). The dashed line
is the approximation from Eq.\ (\ref{lines}) and almost coincides with the curve for $\tilde{\mu}_B=4$. Our model does not 
include a finite pion mass. It can be expected that the effect of the pion mass shifts the curves such that they start at $\mu_I=\pm m_\pi$ 
(corresponding to $\tilde{\mu}_I\simeq \pm 0.375$
with the choice (\ref{fitparams})) instead of at $\mu_I=0$.} 
}

Let us now discuss the resulting phase diagram in the $\mu_B$-$\mu_I$ plane. Firstly, we consider the case of 
a vanishing magnetic field, $H_{\rm em}=0$.
From Eq.\ (\ref{DGapprox}) we see that, in this case, the pion phase is favored in the entire $\mu_B$-$\mu_I$ plane
except for the $\mu_B$ axis. To understand this result we recall several features of our model. 
We treat the fermions as massless (in most applications of the Sakai-Sugimoto model, this 
approximation is used; for a discussion about how to incorporate finite mass effects into the model see Ref.\ \cite{McNees:2008km}).
Therefore, a charged pion condensate appears for arbitrarily small isospin chemical potentials and not only beyond a finite threshold given 
by the pion mass. Moreover, since
we consider the confined phase, we cannot account for phases where there is a vanishing $\langle\bar{u}u\rangle$ and a nonvanishing 
$\langle\bar{d}d\rangle$ condensate or vice versa. In other words, we cannot connect the up-flavor branes and leave the down-flavor 
branes disconnected, as done in Ref.\ \cite{Parnachev:2007bc}, where the deconfined (but chirally broken) phase was considered.
And finally, in our approach we do not see a phase transition 
to the chirally restored phase. Since we are in the confined phase, where the subspace of the 
compactified extra dimension $x_4$ and the 
holographic coordinate $z$ is cigar-shaped, the D8 and $\overline{\rm D8}$ branes {\it must} connect, i.e., chiral symmetry must be
broken for all values of the chemical potentials. The chiral symmetry can only be restored above the deconfinement phase transition at 
$T_c = M_{\rm KK}/(2\pi)$, i.e., using (\ref{fitparams}), $T_c\approx 150 \, {\rm MeV}$.
Taking into account these restrictions, our phase diagram at vanishing magnetic field is in accordance for instance with 
Refs.\ \cite{Parnachev:2007bc,Boer:2008ct,He:2005nk}.

\FIGURE[t]{\label{figphasediag}
{ \centerline{\def\epsfsize#1#2{0.7#1}
\includegraphics[width=0.5\textwidth]{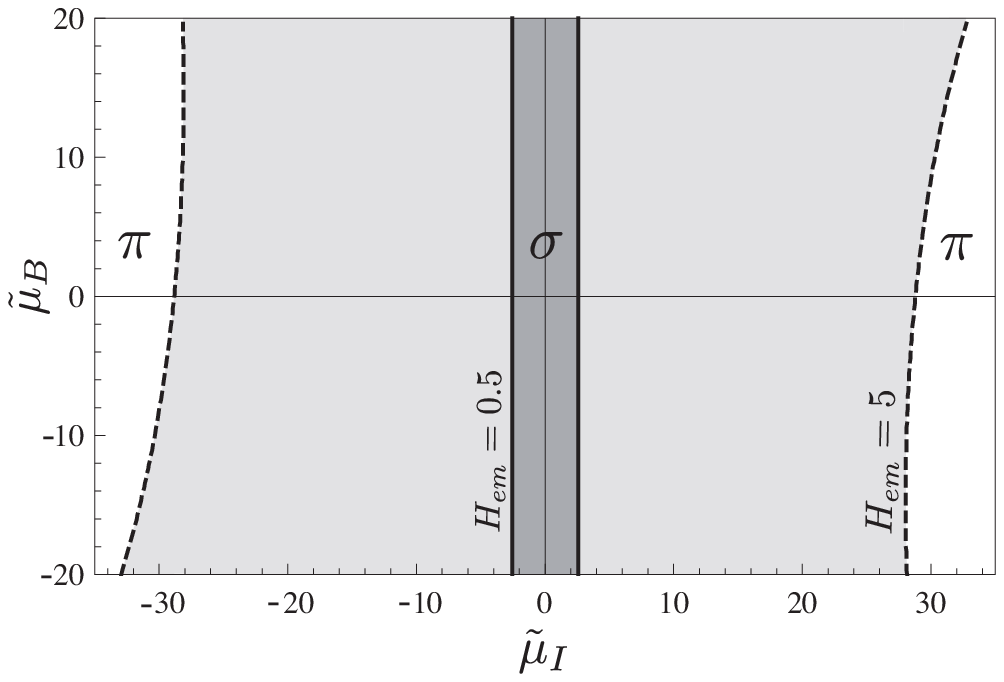}}
}
\caption{Phase diagram for the $\sigma$ and $\pi$ phases in the plane of baryon and isospin chemical potentials. We have chosen 
two different values of the dimensionless magnetic field, \mbox{$H_{\rm em}=0.5$} (solid lines, dark shaded sigma phase) and 
$H_{\rm em}=5$ (dashed lines, light shaded sigma phase). The sigma phase contains meson supercurrents while the pion phase
contains an isotropic $\pi^\pm$ condensate and exhibits the Meissner effect. All lines indicate first order phase 
transitions. The units of this plot, upon fitting the parameters of the model according to (\ref{fitparams}), 
are $\tilde{\mu}_{B,I}\approx \mu_{B,I}/(370\,{\rm MeV})$ and $H_{\rm em}\approx {\cal H}_{\rm em}/(1.8\cdot 10^{19}\,{\rm G})$. Hence, 
due to the huge scales, even compared to magnetar scales, this phase
diagram is rather of academic interest; for more realistic chemical potentials and magnetic fields, the simple approximation (\ref{lines}) is 
sufficient.  
}
}

Next, we discuss the case of a nonzero magnetic field. The phase 
diagram for two different magnetic fields is shown in Fig.\ \ref{figphasediag}. From 
Eq.\ (\ref{lines}) we see that a region for the sigma phase opens up, with  
straight phase transition lines independent of $\mu_B$. These lines start to bend for larger magnetic fields. We may use the numerical 
values of the parameters of the model given in Eq.\ (\ref{fitparams}) and below that equation for some (very) rough quantitative 
predictions from this phase diagram.
First we notice that a dimensionless field $H_{\rm em}=1$ corresponds
to\footnote{In the natural Heavyside-Lorentz system of units of
particle physics, a magnetic field strength of 1~eV$^2$ corresponds to
51.189\ldots G in the Gaussian system.}
${\cal H}_{\rm em}\simeq 2\cdot 10^{19} \, {\rm G}$, about 4 -- 5 orders of magnitude larger than the surface field of magnetars, and 
most likely even larger than the magnetic field in the interior of the star.
For the chemical potentials we find that $\tilde{\mu}_{B,I}=1$ corresponds to \mbox{$\mu_{B,I}\simeq 400\, {\rm MeV}$}. 
As a comparison, a typical baryon chemical potential for neutron stars is 
at most \mbox{$\mu_B\lesssim 1500\, {\rm MeV}$}, corresponding to $\tilde{\mu}_B\simeq4$.
Now, as a rough estimate, let us assume an isospin chemical potential of 1/10 times the baryon chemical potential
in a neutron star, i.e., $\tilde{\mu}_I\simeq 0.4$. Then, the phase transition from the charged pion phase to the 
sigma phase occurs at a very large magnetic field of approximately ${\cal H}_{\rm em} \simeq 1.6 \cdot 10^{18} \, {\rm G}$. 
In other words, the 
charged pion condensate at finite isospin chemical potentials appears to be, in terms of realistic values for the magnetic field, very robust. 

The superconducting properties of a charged pion condensate have been studied
in conventional
chiral models \cite{Harrington:1977py,Harrington:1978gn,Anisimov:1979kt,Migdal:1990vm}. 
There,
%\footnote{Although not so important for
%the orders-of-magnitude estimates considered here, 
%it should be noted that in the
%conversion into Gaussian units in Refs.~\cite{Harrington:1977py,Harrington:1978gn} 
%a factor $\sqrt{4\pi}$, i.e., half an order of magnitude, was missed.}
for an isotropic charged pion condensate, 
the scale of the critical magnetic field is set 
by $m_{\pi}^2$, which is of the order of $10^{18} \, {\rm G}$,
whereas an anisotropic charged pion condensate has been argued to
give a critical magnetic field of the order of $10^{19} \, {\rm G}$.
In our model, in the $\pi$ phase we have
an isotropic $\pi$ condensate and a vanishing $m_\pi$ (see in this
context Ref.~\cite{Abuki:2008wm}), 
but nevertheless we have obtained
a critical field of comparable magnitude.
%, even somewhat larger.
%A trivial but 
%important point in realistic applications for neutron stars is of course
%that with a nonvanishing pion mass the charged pion condensate is
%always absent for $|\mu_I|<m_\pi$ and therefore the critical lines
%in Fig.~\ref{figHcrit} should presumably
%move apart to start at values corresponding
%to $\pm m_\pi$ (at approximately $\pm 0.375$ for $\tilde\mu_I$
%for the parameters we
%have chosen following \cite{Sakai:2004cn,Sakai:2005yt,Hashimoto:2008zw}),
%once finite quark masses are included (cf.\ Ref.~\cite{McNees:2008km}).
It should be noted that in conventional chiral models
the charged pion condensate has been found to behave as a type-II superconductor \cite{Harrington:1977py,Harrington:1978gn,Anisimov:1979kt},
which means that there is another, smaller critical field strength, above which
the magnetic field can penetrate in the form of magnetic flux tubes.
By considering only homogeneous fields, we have of course not taken
this possibility into account. 

%%%%%%%%%%%%%%%%%%%%%%%%%%%%%%%%%%%%%%%%%%%%%%%%%%%%%%%%%%%%%%%%%%%%%%%%%%%%%%%%%%%%%%%%%%%%
\section{Summary and outlook}
\label{sec:summary}
%%%%%%%%%%%%%%%%%%%%%%%%%%%%%%%%%%%%%%%%%%%%%%%%%%%%%%%%%%%%%%%%%%%%%%%%%%%%%%%%%%%%%%%%%%%%

We have employed the Sakai-Sugimoto model to study chiral symmetry breaking in a two-flavor system with a magnetic field and baryon and 
isospin chemical potentials. The model consists of 
a D4/D8-$\overline{\rm D8}$ system in which a bulk gauge symmetry on the D8 and $\overline{\rm D8}$ branes corresponds to a global 
(flavor) symmetry at the boundary which is interpreted as chiral symmetry. Here, left- and right-handed fermions are separated by a fifth 
extra dimension. The electromagnetic subgroup of this flavor symmetry group has been used to incorporate an external magnetic field. 
Our results are independent of temperature, valid for temperatures below the critical temperature for chiral symmetry breaking.
We have discussed how the model can account for different Goldstone boson condensates, two of which we have described within 
our ansatz of abelian gauge field components. Starting from the D-brane action, consisting of Yang-Mills plus Chern-Simons contributions,
we have solved the equations of motion analytically and computed the Gibbs free energies of these two phases. 

The first phase, briefly termed sigma phase, is a linear combination of the usual chiral sigma condensate and a neutral 
pion condensate. We have found that for nonzero magnetic fields this phase exhibits nonzero meson supercurrents. 
In addition, normal currents are generated. For both types of currents, there exist counter-propagating
currents such that the net vector current of the system is zero. This is reminiscent of unconventional (anisotropic) superfluids and superconductors
in condensed matter systems or deconfined dense quark matter. We have found
that the baryon and isospin densities in the system, which obviously vanish without a magnetic field, become nonzero once a magnetic
field is switched on. As a consequence, also an electric charge appears in the sigma phase. 

In the second phase, briefly termed pion phase, charged pions form a condensate. This phase reacts very differently to a magnetic field.
It acts as an electromagnetic superconductor, and thus expels the magnetic field due to the Meissner effect. We have seen that the 
assumption of a nonzero magnetic field would lead to infrared divergences in the energy density which cannot be removed by holographic 
renormalization. Therefore, we have introduced
supercurrents which induce a magnetic field opposite, but equal in magnitude, to the externally applied field. Then the 
total magnetic field in the $\pi$ phase vanishes, and a consistent treatment without divergences is possible. In contrast to the $\sigma$
phase, the $\pi$ phase, due to the Meissner effect, appears unaltered under the influence of a magnetic field. In particular, the baryon number
and all currents (except for the supercurrents cancelling the external magnetic field) vanish. 

Besides the calculation of the supercurrents and the observation of the Meissner effect in the charged $\pi$ condensate, the main result
of our work is the free energy comparison between the two phases and the resulting phase diagram in the $\mu_B$-$\mu_I$ plane. 
For a vanishing magnetic field, 
a nonzero isospin chemical potential leads to the rotation of the sigma condensate into a charged pion condensate. This is expected from studies 
using the same and other models \cite{Parnachev:2007bc,Klein:2003fy,Boer:2008ct,Abuki:2008wm,Andersen:2007qv}. In the present study, which does not
include quark masses, this means that 
in the absence of a magnetic field the pion condensate is favored over the sigma condensate in the entire $\mu_B$-$\mu_I$ plane. For a nonzero
magnetic field the rotation is partially undone, i.e., for a given external magnetic field, there is a region for sufficiently 
small $\mu_I$ where the $\sigma$ phase is favored over the $\pi$ phase. This is not unlike the transition in a metal from its superconducting to its
normal conducting state. We have found that for small magnetic fields, the critical magnetic field for this phase transition
is linear in $\mu_I$ and independent of $\mu_B$, $H_c\propto |\mu_I|$. As a quantitative estimate from our result
we have discussed that for magnetic fields on compact star scales, the charged pion phase at nonzero $\mu_I$ is very robust. 
More precisely, magnetic fields of the order of $10^{18}\, {\rm G}$ (well beyond surface magnetic fields of magnetars, although conceivable for their interiors) are needed to induce a 
phase transition from the $\pi$ to the $\sigma$ phase for isospin chemical potentials of the order of $150\, {\rm MeV}$. 

There are several interesting extensions to our work. One could consider the case of 
not maximally separated D8 and $\overline{\rm D8}$ branes. This would allow for the possibility of the chirally restored phase even at small 
temperatures and thus might answer the question of how a magnetic field affects this symmetry restoration. 
This extension requires to solve an additional equation of motion for the geometry of the D8 and $\overline{\rm D8}$ branes, probably 
making an analytical solution impossible. One may also study the question whether the solutions found here (anisotropic
but homogeneous) are stable against formation of crystalline structures. 
Moreover, since charged pion condensates have been found to behave
as type-II superconductors in conventional chiral models \cite{Harrington:1977py,Harrington:1978gn,Anisimov:1979kt}, it
would be interesting to 
consider inhomogeneous vortex-like configurations of magnetic fields.
It is also important to check whether the states we have described are stable with respect to other meson condensates. We already know
that without magnetic field a rho meson condensate is expected to form for sufficiently large isospin chemical potentials 
\cite{Aharony:2007uu}. 
Another problem, 
not directly related to the present study, but intimately related to the interplay of chirality and a magnetic 
field, is the chiral magnetic effect \cite{Fukushima:2008xe}. It seems that the Sakai-Sugimoto model provides all ingredients to
study this effect.

\acknowledgments
We thank M.\ Alford, D.\ Grumiller, E.\ Fraga, C.\ Herzog, 
M.\ Kaminski, K.\ Landsteiner, G.\ Lifschytz, E.\ Lopez, K.\ Peeters, R.\ Rashkov, D.T.\ Son, and A.\ Vuorinen for valuable 
comments and discussions and acknowledge support by the FWF projects P19526, P19958.

\appendix

%%%%%%%%%%%%%%%%%%%%%%%%%%%%%%%%%%%%%%%%%%%%%%%%%%%%%%%%%%%%%%%%%%%%%%%%%%%%%%%%%%%%%%%%%%%%
\section{General form of equations of motion}
\label{Appeqs}
%%%%%%%%%%%%%%%%%%%%%%%%%%%%%%%%%%%%%%%%%%%%%%%%%%%%%%%%%%%%%%%%%%%%%%%%%%%%%%%%%%%%%%%%%%%%

Here we present the general form of the variations of the Yang-Mills Lagrangian ${\cal L}_{\rm YM}$ and the Chern-Simons Lagrangian 
${\cal L}_{\rm CS}$ (which can be read off from Eqs.\ (\ref{SYM}) and (\ref{SCS})) for the chirally broken phase.
The variation of the Yang-Mills Lagrangian is
\be
\frac{\delta {\cal L}_{\rm YM}}{\delta {\cal A}_\lambda^a} = -2T_8V_4(2\pi\alpha')^2
{\rm Tr}\left[(t_a\partial_\rho -i[t_a,{\cal A}_\rho])e^{-\Phi}\sqrt{g}\,g^{\mu\lambda}g^{\rho\sigma}{\cal F}_{\sigma\mu}\right] \, ,
\ee
where the greek indices run over $0,1,2,3,u$, and where $t_0\equiv {\bf 1}/2$, $t_a\equiv \tau_a/2$, 
according to the convention (\ref{convention}). Consequently,
\begin{subequations} \label{dYM1}
\bea
-\frac{3u_{\rm KK}^{3/2}}{4\kappa M_{\rm KK}^2} \frac{\delta {\cal L}_{\rm YM}}{\delta \hat{A}_0}&=&
\partial_u\left(\frac{u^{5/2}f^{1/2}}{v}\hat{F}_{u0}\right)+
\partial_i\left(\frac{R^3v}{u^{1/2}f^{1/2}}\hat{F}_{i0}\right) \, , \\
-\frac{3u_{\rm KK}^{3/2}}{4\kappa M_{\rm KK}^2}\frac{\delta {\cal L}_{\rm YM}}{\delta \hat{A}_i}&=&
\partial_u\left(\frac{u^{5/2}f^{1/2}}{v}\hat{F}_{ui}\right)+
\partial_0\left(\frac{R^3v}{u^{1/2}f^{1/2}}\hat{F}_{0i}\right)+\partial_j\left(\frac{R^3v}{u^{1/2}f^{1/2}}\hat{F}_{ji}\right) \, , \quad\; 
\label{dYM12}\\
-\frac{3u_{\rm KK}^{3/2}}{4\kappa M_{\rm KK}^2}\frac{\delta {\cal L}_{\rm YM}}{\delta \hat{A}_u}&=&
\partial_0\left(\frac{u^{5/2}f^{1/2}}{v}\hat{F}_{0u}\right)+
\partial_i\left(\frac{u^{5/2}f^{1/2}}{v}\hat{F}_{iu}\right) \, , 
\eea
\end{subequations}
and
\begin{subequations} \label{dYM2}
\bea
-\frac{3u_{\rm KK}^{3/2}}{4\kappa M_{\rm KK}^2}\frac{\delta {\cal L}_{\rm YM}}{\delta A_0^a}&=&
(\delta_{ac}\partial_u+A_u^b\epsilon_{abc})\frac{u^{5/2}f^{1/2}}{v}F_{u0}^c+
(\delta_{ac}\partial_i+A_i^b\epsilon_{abc})\frac{R^3v}{u^{1/2}f^{1/2}}F_{i0}^c \, , \quad\; \\
-\frac{3u_{\rm KK}^{3/2}}{4\kappa M_{\rm KK}^2}\frac{\delta {\cal L}_{\rm YM}}{\delta A_i^a}&=&
(\delta_{ac}\partial_u+A_u^b\epsilon_{abc})\frac{u^{5/2}f^{1/2}}{v}F_{ui}^c+
(\delta_{ac}\partial_0+A_0^b\epsilon_{abc})\frac{R^3v}{u^{1/2}f^{1/2}}F_{0i}^c \non
&&+\,
(\delta_{ac}\partial_j+A_j^b\epsilon_{abc})\frac{R^3v}{u^{1/2}f^{1/2}}F_{ji}^c  \, , \label{dCS12} \\
-\frac{3u_{\rm KK}^{3/2}}{4\kappa M_{\rm KK}^2}\frac{\delta {\cal L}_{\rm YM}}{\delta A_u^a}&=&
(\delta_{ac}\partial_0+A_0^b\epsilon_{abc})\frac{u^{5/2}f^{1/2}}{v}F_{0u}^c+
(\delta_{ac}\partial_i+A_i^b\epsilon_{abc})\frac{u^{5/2}f^{1/2}}{v}F_{iu}^c \, , 
\eea
\end{subequations}
where the indices $i,j,k$ run over 1,2,3, and where we used  
\be \label{help3}
\frac{T_8 V_4(2\pi\alpha')^2R^{3/2}}{g_s} = \frac{4\kappa M_{\rm KK}^2}{3u_{\rm KK}^{3/2}} \, , 
\ee
with $\kappa$ defined in Eq.\ (\ref{kappa}).

The variations of the Chern-Simons Lagrangian with respect to the ${\bf 1}$ and $\tau_3$ components are 
\begin{subequations}
\bea
\frac{\delta {\cal L}_{\rm CS}}{\delta \hat{A}_\mu}&=&-i\frac{\kappa\alpha}{4}
(F_{\nu\rho}^a F_{\sigma\lambda}^a+\hat{F}_{\nu\rho}\hat{F}_{\sigma\lambda}) \epsilon^{\mu\nu\rho\sigma\lambda} \, , \\
\frac{\delta {\cal L}_{\rm CS}}{\delta A_\mu^a}&=&-i\frac{\kappa\alpha}{2}
\hat{F}_{\nu\rho} F_{\sigma\lambda}^a \epsilon^{\mu\nu\rho\sigma\lambda}  \, , 
\eea
\end{subequations}
with $\alpha$ defined in Eq.\ (\ref{alpha}). Consequently,
\begin{subequations} \label{dCS1}
\bea
\frac{\delta {\cal L}_{\rm CS}}{\delta \hat{A}_0}&=&i\kappa\alpha(F_{ui}^a F_{jk}^a+\hat{F}_{ui}\hat{F}_{jk})\epsilon^{ijk} \, , \\
\frac{\delta {\cal L}_{\rm CS}}{\delta \hat{A}_i}&=& i\kappa\alpha\left(2F_{j0}^aF_{uk}^a-F_{u0}^aF_{jk}^a+2\hat{F}_{j0}\hat{F}_{uk}-
\hat{F}_{u0}\hat{F}_{jk}\right)\epsilon^{ijk}\, , \\
\frac{\delta {\cal L}_{\rm CS}}{\delta \hat{A}_u}&=& i\kappa\alpha(F_{i0}^a F_{jk}^a+\hat{F}_{i0}\hat{F}_{jk})\epsilon^{ijk} \, , 
\eea
\end{subequations}
and
\begin{subequations} \label{dCS2}
\bea
\frac{\delta {\cal L}_{\rm CS}}{\delta A_0^a}&=& i\kappa\alpha(F_{ui}^a \hat{F}_{jk}+F_{jk}^a \hat{F}_{ui})\epsilon^{ijk}\, , \\
\frac{\delta {\cal L}_{\rm CS}}{\delta A_i^a}&=& i\kappa\alpha\left(2F_{uk}^a\hat{F}_{j0}-F_{jk}^a\hat{F}_{u0}
+2F_{j0}^a\hat{F}_{uk}-F_{u0}^a\hat{F}_{jk}\right)\epsilon^{ijk}\, , \\
\frac{\delta {\cal L}_{\rm CS}}{\delta A_u^a}&=& i\kappa\alpha (F_{i0}^a \hat{F}_{jk}+F_{jk}^a\hat{F}_{i0})\epsilon^{ijk}\, . 
\eea
\end{subequations}
As mentioned in the main part of the paper, 
we consider maximally separated branes $L=\pi/M_{\rm KK}$, for which the embedding of the D8 branes is trivial, 
$\partial_u x_4=0$, and thus $v=1$ (see Eq.\ (\ref{vvtilde})). This simplifies the above expressions and also ensures that there 
is no equation of motion for $x_4(u)$. The expressions (\ref{dYM1}), (\ref{dYM2}), (\ref{dCS1}), (\ref{dCS2}) are used in Sec.\ \ref{sec:eqsmotion}
to derive the field equations for our specific ansatz. 

%%%%%%%%%%%%%%%%%%%%%%%%%%%%%%%%%%%%%%%%%%%%%%%%%%%%%%%%%%%%%%%%%%%%%%%%%%%%%%%%%%%%%%%%%%%%
\section{Solving the equations of motion for constant magnetic fields}
\label{appsigma}
%%%%%%%%%%%%%%%%%%%%%%%%%%%%%%%%%%%%%%%%%%%%%%%%%%%%%%%%%%%%%%%%%%%%%%%%%%%%%%%%%%%%%%%%%%%%

In this appendix we solve the equations of motion for a constant magnetic field for general boundary conditions.
The resulting general expressions are instructive to see the structure and symmetries of the solution. By inserting the boundary conditions 
from Table \ref{tableboundary} into the general expressions  
we obtain the solution for the sigma phase, see 
Eqs.\ (\ref{gaugefields1}) (the pion phase requires a nonconstant magnetic field and is discussed in Appendix \ref{apppi}). 
The general boundary conditions used here are denoted by
\begin{subequations} \label{boundgeneral1}
\bea
\hat{A}_0(\pm\infty) &=& 2\mu_B^{R,L} \, , \qquad A_0(\pm\infty) = 2\mu_I^{R,L} \, , \\
\hat{A}_3(\pm\infty) &=& 2\hat{\jmath}^{R,L} \, , \qquad A_3(\pm\infty) = 2\jmath^{R,L} \, , 
\eea
\end{subequations}
where the upper (lower) sign corresponds to $R$ ($L$). 
It is convenient to express the boundary values in terms of their sums and differences,
\begin{subequations}
\bea
\mu_{B,I}^V&\equiv& \frac{\mu_{B,I}^R+\mu_{B,I}^L}{2} \, , \qquad  \mu_{B,I}^A\equiv \frac{\mu_{B,I}^R-\mu_{B,I}^L}{2} \, ,  \\
\hat{J}&\equiv& \frac{\hat{\jmath}^R+\hat{\jmath}^L}{2} \, , \qquad  \hat{\jmath}\equiv \frac{\hat{\jmath}^R-\hat{\jmath}^L}{2} \, , \qquad
J\equiv \frac{\jmath^R+\jmath^L}{2} \, , \qquad  \jmath\equiv \frac{\jmath^R-\jmath^L}{2} \, .
\eea
\end{subequations}
Here, $V$ and $A$ stand for the vector and axial parts of the chemical potentials. 
%The physical boundary conditions are obtained 
%by setting $\mu_B^A=\mu_I^A=\hat{J}=J=0$.  

The general solution for (\ref{newdiffs}) with the magnetic field (\ref{notwist}) is  
\begin{subequations} \label{solutions}
\bea
F_0^+&=&c_1\zeta_+^{-1}+c_2\zeta_+ \, , \qquad F_0^-=d_1\zeta_-^{-1}+d_2\zeta_- \, , \\
F_3^+&=&-c_1\zeta_+^{-1}+c_2\zeta_+ \, , \qquad F_3^-=-d_1\zeta_-^{-1}+d_2\zeta_- \, , 
\eea
\end{subequations}
with constants $c_1$, $c_2$, $d_1$, $d_2$ and with
\be \label{zetapm}
\zeta_\pm(z)\equiv e^{(\hat{B}\pm B)\arctan z} \, .
\ee
Consequently, from Eqs.\ (\ref{F03}) we obtain
\begin{subequations} \label{fieldstrengths}
\bea
k\hat{F}_{z0}&=&c_1\zeta_+^{-1}+c_2\zeta_++d_1\zeta_-^{-1}+d_2\zeta_- \, , \\
kF_{z0}&=&c_1\zeta_+^{-1}+c_2\zeta_+-d_1\zeta_-^{-1}-d_2\zeta_- \, , \\
k\hat{F}_{z3}&=&-c_1\zeta_+^{-1}+c_2\zeta_+-d_1\zeta_-^{-1}+d_2\zeta_- \, , \\
kF_{z3}&=&-c_1\zeta_+^{-1}+c_2\zeta_++d_1\zeta_-^{-1}-d_2\zeta_- \, .
\eea
\end{subequations}
Here and in the remainder of this and the following appendices we often omit the argument $z$ in the various functions for the sake of brevity.
For the integration of the field strengths we use 
\be
\int dz \frac{\zeta_\pm(z)}{k(z)} = \frac{\zeta_\pm(z)}{\hat{B}\pm B}\, , \qquad  
\int dz \frac{\zeta_\pm^{-1}(z)}{k(z)} = -\frac{\zeta_\pm^{-1}(z)}{\hat{B}\pm B} \, .
\ee
This yields the gauge fields
\begin{subequations} \label{fields}
\bea
\hat{A}_0&=& -\frac{c_1\zeta_+^{-1}}{\hat{B}+B}+\frac{c_2\zeta_+}{\hat{B}+B}-\frac{d_1\zeta_-^{-1}}{\hat{B}-B}+\frac{d_2\zeta_-}{\hat{B}-B}+
\hat{a}_0 \, , \\
A_0&=& -\frac{c_1\zeta_+^{-1}}{\hat{B}+B}+\frac{c_2\zeta_+}{\hat{B}+B}+\frac{d_1\zeta_-^{-1}}{\hat{B}-B}-\frac{d_2\zeta_-}{\hat{B}-B} +a_0\, , \\
\hat{A}_3&=& \frac{c_1\zeta_+^{-1}}{\hat{B}+B}+\frac{c_2\zeta_+}{\hat{B}+B}+\frac{d_1\zeta_-^{-1}}{\hat{B}-B}+\frac{d_2\zeta_-}{\hat{B}-B}+
\hat{a}_3 \, , \\
A_3&=& \frac{c_1\zeta_+^{-1}}{\hat{B}+B}+\frac{c_2\zeta_+}{\hat{B}+B}-\frac{d_1\zeta_-^{-1}}{\hat{B}-B}-\frac{d_2\zeta_-}{\hat{B}-B}+a_3 \, ,
\eea
\end{subequations}
with integration constants $\hat{a}_0$, $a_0$, $\hat{a}_3$, $a_3$. We determine the eight constants from the eight boundary conditions
(\ref{boundgeneral1}). This yields the gauge fields 
\begin{subequations} \label{gaugefields0}
\bea
\hat{A}_0 &=& 2\mu_B^V + \mu_B^A(S_++S_-)+\mu_I^A(S_+-S_-) \non
&&+\,\hat{\jmath}(C_++C_--T_+)+\jmath(C_+-C_--T_-) 
\, ,  \label{hattemp} \\
A_0 &=& 2\mu_I^V +\mu_I^A(S_++S_-) + \mu_B^A(S_+-S_-) \non
&&+\,\jmath(C_++C_--T_+) +\hat{\jmath}(C_+-C_--T_-) \, , \label{temp} \\
\hat{A}_3 &=& 2\hat{J} + \hat{\jmath}(S_++S_-)+\jmath(S_+-S_-) \non
&&+\,\mu_B^A(C_++C_--T_+)+\mu_I^A(C_+-C_--T_-) \, , \label{hatspat} \\
A_3 &=& 2J +\jmath(S_++S_-)+ \hat{\jmath}(S_+-S_-) \non
&&+\, \mu_I^A(C_++C_--T_+) +\mu_B^A(C_+-C_--T_-) \, , \label{spat}
\eea
\end{subequations}
and ($k(z)$ times) the field strengths
\begin{subequations} \label{kF}
\bea
k\hat{F}_{z0} &=& \mu_B^A[\hat{B}(C_++C_-)+B(C_+-C_-)]+\mu_I^A[\hat{B}(C_+-C_-)+B(C_++C_-)] \non
&&+\,\hat{\jmath}[\hat{B}(S_++S_-)+B(S_+-S_-)]+\jmath[\hat{B}(S_+-S_-)+B(S_++S_-)] \, , \\
kF_{z0} &=& \mu_B^A[\hat{B}(C_+-C_-)+B(C_++C_-)]+\mu_I^A[\hat{B}(C_++C_-)+B(C_+-C_-)] \non
&&+\,\hat{\jmath}[\hat{B}(S_+-S_-)+B(S_++S_-)]+\jmath[\hat{B}(S_++S_-)+B(S_+-S_-)] \, , \\
k\hat{F}_{z3} &=& \hat{\jmath}[\hat{B}(C_++C_-)+B(C_+-C_-)]+\jmath[\hat{B}(C_+-C_-)+B(C_++C_-)] \non
&&+\,\mu_B^A[\hat{B}(S_++S_-)+B(S_+-S_-)]+\mu_I^A[\hat{B}(S_+-S_-)+B(S_++S_-)] \, , \\
kF_{z3}&=& \hat{\jmath}[\hat{B}(C_+-C_-)+B(C_++C_-)]+\jmath[\hat{B}(C_++C_-)+B(C_+-C_-)] \non
&&+\,\mu_B^A[\hat{B}(S_+-S_-)+B(S_++S_-)]+\mu_I^A[\hat{B}(S_++S_-)+B(S_+-S_-)] \, , 
\eea
\end{subequations}
where the functions $C_\pm(z)$, $S_\pm(z)$, and $T_\pm$ are defined in Eqs.\ (\ref{CST}). As it should be, 
the gauge fields (\ref{gaugefields0}) transform as a vector under a parity transformation. 
This can be seen as follows. A parity transformation is given by 
$(x_1,x_2,x_3,z)\to(-x_1,-x_2,-x_3,-z)$. In particular, the transformation $z\to -z$ implies a chirality transformation $R\to L$ since
the two halves of the D8/$\overline{\rm D8}$ branes, namely $z>0$ and $z<0$, correspond to right- and left-handed fermions. 
Consequently, a parity transformation acts as $C_\pm(z)\to +C_\pm(z)$, $S_\pm(z)\to -S_\pm(z)$ (since the magnetic fields $\hat{B}$, $B$
are even under parity). For the supercurrents we have $\hat{\jmath},\jmath\to +\hat{\jmath},+\jmath$ and $\hat{J},J\to -\hat{J},-J$. 
Here we have used that the 
Goldstone bosons are pseudoscalars (for a detailed discussion of the parity of the mesons in the Sakai-Sugimoto model see Ref.\ \cite{Sakai:2004cn}).
As a result we see that the temporal components (\ref{hattemp}), (\ref{temp}) have even parity, while the spatial components
(\ref{hatspat}), (\ref{spat}) have odd parity. This statement is true for arbitrary values of the currents $\hat{\jmath},\jmath,\hat{J},J$.
We shall see below that in the case of a charged pion condensate the requirement of a well-defined parity results in conditions for
the supercurrents, see discussion below Eq.\ (\ref{Omega7}).

In order to compute the free energy we note that
\be
C_\pm^2-S_\pm^2=\frac{1}{\sinh^2[\pi(\hat{B}\pm B)/2]} \, .
\ee
Therefore, the following combination of field strengths, needed for the free energy, becomes independent of $z$,
\bea \label{help1}
k^2\left(-\hat{F}_{z0}^2-F_{z0}^2+\hat{F}_{z3}^2+F_{z3}^2\right)
&=&2[(\hat{\jmath}+\jmath)^2-(\mu_B^A+\mu_I^A)^2]\frac{(\hat{B}+B)^2}{\sinh^2[\pi(\hat{B}+ B)/2]} \non
&& \hspace{-1cm}+\,2[(\hat{\jmath}-\jmath)^2-(\mu_B^A-\mu_I^A)^2]\frac{(\hat{B}-B)^2}{\sinh^2[\pi(\hat{B}-B)/2]}
\, .
\eea
Next we use the fact that $S_\pm$ and $C_\pm$ are antisymmetric and symmetric in $z$, respectively, as well as
\be
C_\pm(\infty)=\coth\frac{\pi(\hat{B}\pm B)}{2} \, , \qquad S_\pm(\infty) = 1  \, , 
\ee
to obtain
\bea \label{help2}
&&\left(\hat{A}_0k\hat{F}_{z0}+A_0kF_{z0}-\hat{A}_3k\hat{F}_{z3}-A_3kF_{z3}\right)_{z=-\infty}^{z=\infty}\non
&&=8\mu_B^V(\hat{\jmath}\hat{B}+\jmath B)+8\mu_I^V(\hat{\jmath}B+\jmath\hat{B})-8\hat{J}(\mu_B^A\hat{B}+\mu_I^A B)-\,8J(\mu_B^AB+\mu_I^A\hat{B})\non
&&+\,4[(\mu_B^A+\mu_I^A)^2-(\hat{\jmath}+\jmath)^2](\hat{B}+B)\coth\frac{\pi(\hat{B}+B)}{2} \non
&&+\,4[(\mu_B^A-\mu_I^A)^2-(\hat{\jmath}-\jmath)^2](\hat{B}-B)\coth\frac{\pi(\hat{B}- B)}{2} \, . 
\eea
Inserting Eqs.\ (\ref{help1}) and (\ref{help2}) into Eq.\ (\ref{Omegaconf}) yields the free energy 
\bea \label{Omegasigma3}
\Omega &=& \frac{2\kappa M_{\rm KK}^2}{3}\Big\{\left[(\jmath+\hat{\jmath})^2-(\mu_B^A+\mu_I^A)^2\right]\rho_++
\left[(\jmath-\hat{\jmath})^2-(\mu_B^A-\mu_I^A)^2\right]\rho_- \non
&&-\,4\mu_B^V(\hat{\jmath}\hat{B}+\jmath B)-4\mu_I^V(\hat{\jmath}B+\jmath\hat{B}) \non 
&&+\,4\hat{J}(\mu_B^A\hat{B}+\mu_I^A B)+4J(\mu_B^AB+\mu_I^A\hat{B})\Big\}
\, ,
\eea
with $\rho_\pm$ defined in Eq.\ (\ref{rhopm}). For the behavior of $\rho_\pm$ for small and large magnetic fields see 
Table \ref{tablerhoeta}. We see that if we allowed for nonzero axial chemical potentials $\mu_B^A$, $\mu_I^A$, 
the free energy would be unbounded from below in the directions of the sum of left- and right-handed
supercurrents $\hat{J}$ and $J$. However, in the physical case of the $\sigma$ phase where $\mu_B^A=\mu_I^A=0$ the free 
energy remains bounded and independent of $\hat{J}$ and $J$. The latter is a manifestation of a residual gauge symmetry 
(``residual'' since we have already employed the gauge ${\cal A}_z=0$), i.e., we can choose a gauge where $\hat{J}=J=0$. This is in 
accordance with the discussion in Ref.\ \cite{Sakai:2004cn}, see in particular Eq.\ (5.23) in this reference. 

\TABLE[t]{
\begin{tabular}{|c||c||c|c|} 
\hline
 & small $\hat{B},B$ & \multicolumn{2}{c}{large $|\hat{B}|,|B|$}\vline \\ \hline 
 & & $\;\;|\hat{B}|>|B|\;\;$ & $\;\;|\hat{B}|<|B|\;\;$ \\ \hline\hline
\rule[-1.5ex]{0em}{6ex}$\;\;\rho_\pm\;\;$ & $\displaystyle{\frac{6}{\pi}+\frac{\pi(\hat{B}\pm B)^2}{6}}$ & 
\multicolumn{2}{c}{$2|\hat{B}\pm B|$}\vline \\[2ex] \hline
\rule[-1.5ex]{0em}{6ex}$\rho$ & $\;\;\displaystyle{\frac{12}{\pi}+\frac{5\hat{B}^2+B^2}{15}\pi}\;\;$ & $4|\hat{B}|$ & $2(|\hat{B}|+|B|)$ \\[2ex] \hline
\rule[-1.5ex]{0em}{6ex}$\eta_+$ & $\displaystyle{\frac{\pi\hat{B}B}{3}}$ & $2B\,{\rm sgn}\,\hat{B}$ & 
$\;\;B\,{\rm sgn}\,\hat{B}+\hat{B}\,{\rm sgn}\,B\;\;$ \\[2ex] \hline
$\eta_-$ & $2\hat{B}$ & $2\hat{B}$ & $(|\hat{B}|+|B|)\,{\rm sgn}\,\hat{B}$ \\ \hline
\end{tabular}
\caption{Behavior of the functions $\rho$, $\rho_\pm$, $\eta_\pm$, defined in Eqs.\ (\ref{rhopm}), (\ref{rhoeta}) 
for small and large magnetic fields $\hat{B}$, $B$. We have kept relative magnitude 
and sign of baryon and isospin components arbitrary. They can then later be inserted according to the electric charges of the quarks.
We show the behavior for small magnetic fields up to second order and the behavior for large magnetic fields in leading linear order.
}
\label{tablerhoeta}
}

Minimization of  $\Omega$ with respect to the currents $\hat{\jmath}$, $\jmath$ yields
\begin{subequations}
\bea
\hat{\jmath}&=&\frac{\mu_B^V+\mu_I^V}{2}\frac{\hat{B}+B}{\rho_+}+\frac{\mu_B^V-\mu_I^V}{2}\frac{\hat{B}-B}{\rho_-} \, , \\
\jmath&=&\frac{\mu_B^V+\mu_I^V}{2}\frac{\hat{B}+B}{\rho_+}-\frac{\mu_B^V-\mu_I^V}{2}\frac{\hat{B}-B}{\rho_-} \, ,
\eea
\end{subequations}
and the minimum of the free energy becomes (with $\mu_B^A=\mu_I^A=0$)
\bea \label{Omega0sigma1}
\Omega_0 &=& -\frac{2\kappa M_{\rm KK}^2}{3}
\left[(\mu_B^V+\mu_I^V)^2\frac{(\hat{B}+B)^2}{\rho_+}+(\mu_B^V-\mu_I^V)^2\frac{(\hat{B}-B)^2}{\rho_-}\right] 
%\non
%&&+\,(\mu_B^A+\mu_I^A)^2\rho_+
%+(\mu_B^A-\mu_I^A)^2\rho_-\Big]
\, .
\eea
%We confirm from this form of the free energy what we have claimed above Eq.\ (\ref{casei}): the phase where both left- and right-handed
%isospin components flip their sign is equivalent to the phase considered here. To see this, one simply has to replace $\mu_I^V$, $\mu_I^A$, 
%and $B$ by their negatives. 

%%%%%%%%%%%%%%%%%%%%%%%%%%%%%%%%%%%%%%%%%%%%%%%%%%%%%%%%%%%%%%%%%%%%%%%%%%%%%%%%%%%%%%%%%%%%
\section{Solving the equations of motion for nonconstant magnetic fields}
\label{apppi}
%%%%%%%%%%%%%%%%%%%%%%%%%%%%%%%%%%%%%%%%%%%%%%%%%%%%%%%%%%%%%%%%%%%%%%%%%%%%%%%%%%%%%%%%%%%%

In this appendix we present the general solution to the differential equations (\ref{newdiffs}) for the case of a nonconstant isospin 
magnetic field given in Eq.\ (\ref{twist}). The general expressions given below reduce to the results for the charged pion phase upon inserting
the specific boundary conditions from the second row of Table \ref{tableboundary}. 
The general boundary conditions considered here are the same as the ones given in Eqs.\ (\ref{boundgeneral1}).

Then, the solution of (\ref{newdiffs}) has the same form as given in Eqs.\ (\ref{solutions}) and (\ref{fieldstrengths}), 
with $\zeta_\pm(z)$ replaced by 
\be \label{zetapm1}
\tilde{\zeta}_\pm(z)\equiv e^{(\hat{B}\pm \frac{B}{\pi}\arctan z)\arctan z} \, .
\ee
To obtain the gauge fields we need 
\begin{subequations}
\bea
\int dz \frac{\tilde{\zeta}_+(z)}{k(z)} &=& P_+(z) \, , \qquad \int dz \frac{\tilde{\zeta}_-^{-1}(z)}{k(z)} = -P_-(z)\, , \\  
\int dz \frac{\tilde{\zeta}_+^{-1}(z)}{k(z)} &=& Q_+(z) \, , \qquad \int dz \frac{\tilde{\zeta}_-(z)}{k(z)} = -Q_-(z)\, ,   
\eea
\end{subequations}
with $P_\pm$, $Q_\pm$ given in Eqs.\ (\ref{PQ}).
We shall denote $Q_\pm^+\equiv Q_\pm(+\infty)$, $Q_\pm^-\equiv Q_\pm(-\infty)$, $P_\pm^+\equiv P_\pm(+\infty)$, $P_\pm^-\equiv P_\pm(-\infty)$,
and use $P_-^\pm=P_+^\mp$, $Q_-^\pm=Q_+^\mp$. Hence we can express the values of $P_-$, $Q_-$ at $z=\pm\infty$ through the values of  
$P_+$, $Q_+$ at $z=\mp\infty$. Then, after determining the integration constants from the boundary conditions we can write the gauge fields as 
\begin{subequations} \label{gaugefields01}
\bea
\hat{A}_0 &=& 2\mu_B^V + \mu_B^A(\tilde{S}_++\tilde{S}_-)+\jmath(\tilde{S}_+-\tilde{S}_-) \non 
&&+\, \mu_I^A(\tilde{C}_++\tilde{C}_--\tilde{T}_+) 
+\hat{\jmath}(\tilde{C}_+-\tilde{C}_--\tilde{T}_-) \, , \\
A_0 &=& 2\mu_I^V +\mu_I^A(\tilde{S}_++\tilde{S}_-) +\hat{\jmath}(\tilde{S}_+-\tilde{S}_-) \non 
&&+\, \mu_B^A(\tilde{C}_++\tilde{C}_--\tilde{T}_+)
+\jmath(\tilde{C}_+-\tilde{C}_--\tilde{T}_-) \, , \\
\hat{A}_3 &=& 2\hat{J} + \hat{\jmath}(\tilde{S}_++\tilde{S}_-)+\mu_I^A(\tilde{S}_+-\tilde{S}_-) \non 
&&+\,\jmath(\tilde{C}_++\tilde{C}_--\tilde{T}_+)
+\mu_B^A(\tilde{C}_+-\tilde{C}_--\tilde{T}_-) \, , \\
A_3 &=& 2J + \jmath(\tilde{S}_++\tilde{S}_-)+\mu_B^A(\tilde{S}_+-\tilde{S}_-) \non
&&+\,\hat{\jmath}(\tilde{C}_++\tilde{C}_--\tilde{T}_+)
+\mu_I^A(\tilde{C}_+-\tilde{C}_--\tilde{T}_-) \, , 
\eea
\end{subequations}
and the field strengths as
\begin{subequations}
\bea
k\hat{F}_{z0} &=& \mu_B^A(c_++c_-)+\jmath(c_+-c_-)+\mu_I^A(s_++s_-)+\hat{\jmath}(s_+-s_-) \, , \\
kF_{z0} &=& \mu_I^A(c_++c_-)+\hat{\jmath}(c_+-c_-)+\mu_B^A(s_++s_-)+\jmath(s_+-s_-) \, , \\
k\hat{F}_{z3} &=& \hat{\jmath}(c_++c_-)+\mu_I^A(c_+-c_-)+\jmath(s_++s_-)+\mu_B^A(s_+-s_-) \, , \\
kF_{z3} &=& \jmath(c_++c_-)+\mu_B^A(c_+-c_-)+\hat{\jmath}(s_++s_-)+\mu_I^A(s_+-s_-) \, , 
\eea
\end{subequations}
where $\tilde{C}_\pm$, $\tilde{S}_\pm$, and $\tilde{T}_\pm$ are defined in Eqs.\ (\ref{CSTtilde}), and where 
\begin{subequations}
\bea
c_+(z)&\equiv& \frac{\tilde{\zeta}_+(z)+\tilde{\zeta}_-^{-1}(z)}{P_+^+-P_+^-} \, , \qquad c_-(z)\equiv \frac{\tilde{\zeta}_+^{-1}(z)+\tilde{\zeta}_-(z)}{Q_+^+-Q_+^-} \, , \\
s_+(z)&\equiv& \frac{\tilde{\zeta}_+(z)-\tilde{\zeta}_-^{-1}(z)}{P_+^+-P_+^-} \, , \qquad s_-(z)\equiv \frac{\tilde{\zeta}_+^{-1}(z)-\tilde{\zeta}_-(z)}{Q_+^+-Q_+^-} \, .
\eea
\end{subequations}
(These additional definitions were not necessary in the case of constant magnetic fields, 
since there the integration of the solution could be expressed in terms of the same functions as the solution itself.)

We now have to check the behavior of the gauge fields (\ref{gaugefields01}) under a parity transformation. For the pion phase we
have $\mu_B^A=\mu_I^V=0$. We have to require $\hat{A}_0\to +\hat{A}_0$, $A_0\to -A_0$, $\hat{A}_3\to -\hat{A}_3$, $A_3 \to +A_3$ (note
the additional ``twist'' for the isospin components originating from the isospin rotation explained in Sec.\ \ref{sec:rotate}). Since 
$\tilde{C}_\pm(z)\to +\tilde{C}_\pm(z)$, $\tilde{S}_\pm(z)\to -\tilde{S}_\pm(z)$, and 
$\hat{\jmath},\jmath\to +\hat{\jmath},+\jmath$ and $\hat{J},J\to -\hat{J},-J$ under a parity transformation, we have to require 
\be \label{JIjI}
J = \jmath = 0 \, .
\ee
We shall continue with the general solution but have to keep this condition in mind for the final result.

For the free energy we first note that the following combinations are independent of $z$, 
\be
c_+c_-+s_+s_-=\frac{4}{(P_+^+-P_+^-)(Q_+^+-Q_+^-)} \, , \qquad s_+c_-+s_-c_+ = 0  \, .
\ee
Then, we find 
\bea
k^2\left(-\hat{F}_{z0}^2-F_{z0}^2+\hat{F}_{z3}^2+F_{z3}^2\right) &=& 
16\frac{(\hat{\jmath}^2+\jmath^2)-[(\mu_B^A)^2+(\mu_I^A)^2]}{(P_+^+-P_+^-)(Q_+^+-Q_+^-)} \, .
\eea
Next we use that $c_\pm$ and $s_\pm$ are symmetric and antisymmetric in $z$, respectively, and denote $c_\pm^+\equiv c_\pm(\infty)=c_\pm(-\infty)$,
$s_\pm^+\equiv s_\pm(\infty)=-s_\pm(-\infty)$. Then,
\bea
&&\left(\hat{A}_0k\hat{F}_{z0}+A_0kF_{z0}-\hat{A}_3k\hat{F}_{z3}-A_3kF_{z3}\right)_{z=-\infty}^{z=\infty}\non
&&=4(s_+^++s_-^+)(\mu_B^V\mu_I^A+\mu_I^V\mu_B^A-\jmath \hat{J}-\hat{\jmath}J)
+4(s_+^+-s_-^+)(\hat{\jmath}\mu_B^V+\jmath\mu_I^V-\hat{J}\mu_B^A-J\mu_I^A)\non
&&+4(c_+^++c_-^+)[(\mu_B^A)^2+(\mu_I^A)^2-(\hat{\jmath}^2+\jmath^2)] \, .
\eea
Inserting this into the free energy (\ref{Omegaconf}) yields
\bea \label{Omega7}
\Omega &=& \frac{2\kappa M_{\rm KK}^2}{3}\Big\{\left[\hat{\jmath}^2+\jmath^2-(\mu_B^A)^2-(\mu_I^A)^2\right]\,\rho 
-2(\mu_B^V\mu_I^A+\mu_I^V\mu_B^A)\,\eta_+-2(\mu_B^V\hat{\jmath}+\mu_I^V\jmath)\,\eta_- \non
&&+\,2\hat{J}(\mu_B^A\eta_-+\jmath\eta_+)+2J(\mu_I^A\eta_-+\hat{\jmath}\eta_+)\Big\} \, ,
\eea
with $\rho$ and $\eta_\pm\equiv s_+^+\pm s_-^+$ given in Eqs.\ (\ref{rhoeta}); their behavior for small and large magnetic fields can be found in 
Table \ref{tablerhoeta}. As in the case of constant magnetic fields discussed in the previous appendix, see Eq.\ (\ref{Omegasigma3}), the 
free energy is unbounded from below without further constraints. This can be seen by computing 
the matrix of second derivatives $\partial^2\Omega/(\partial x_m\partial x_n)$ with $x_m,x_n=\hat{\jmath},\jmath,\hat{J},J$. This matrix 
has eigenvalues $2\kappa M_{\rm KK}^2/3\,[\rho\pm(\rho^2+4\eta_+^2)^{1/2}]$, two of which are negative for all magnetic fields.
However, we already know from the requirement of a well-defined parity of the gauge fields that $J=\jmath=0$. Then, with 
$\mu_B^A=\mu_I^V=0$, as required for the charged pion condensate, we see that the free energy becomes bounded from below. The only remaining 
supercurrent with 
respect to which we need to minimize the free energy is then $\hat{\jmath}$. The sum of left- and right-handed supercurrents, $\hat{J}$, remains 
undetermined, which is, as mentioned for the case of the sigma phase below Eq.\ (\ref{Omegasigma3}), 
a consequence of the residual gauge freedom. We may thus set $\hat{J}=0$.

We can now minimize with respect to $\hat{\jmath}$,
\be \label{jI}
\hat{\jmath}=\mu_B^V\frac{\eta_-}{\rho} \, , %\qquad \jmath=\mu_I^V\frac{\eta_+}{\rho} \, .
\ee
and insert this back into the free energy,
\be
\Omega_0 = -\frac{2\kappa M_{\rm KK}^2}{3} \left\{(\mu_B^V)^2\frac{\eta_-^2}{\rho}+(\mu_I^A)^2\rho
+2\eta_+ \mu_B^V\mu_I^A \right\} \, .
\ee 

%%%%%%%%%%%%%%%%%%%%%%%%%%%%%%%%%%%%%%%%%%%%%%%%%%%%%%%%%%%%%%%%%%%%%%%%%%%%%%%%%%%%%%%%%%%%
\section{Equations of motion and free energy in the chirally restored phase}
\label{sec:restored}
%%%%%%%%%%%%%%%%%%%%%%%%%%%%%%%%%%%%%%%%%%%%%%%%%%%%%%%%%%%%%%%%%%%%%%%%%%%%%%%%%%%%%%%%%%%%

Within our approximation of treating the flavor branes as probe branes, 
the free energies discussed in the main part of the paper are negligible for the finite-temperature phase transition to the 
chirally restored phase. It is rather the background geometry which is responsible for this phase transition 
\cite{Witten:1998zw,Horigome:2006xu}. Therefore, our approach 
cannot show magnetic-field induced corrections beyond the order of $N_f/N_c$ to the critical temperature $T_c$ for chiral symmetry breaking. This
is different when the D8 and $\overline{\rm D8}$ branes are not maximally separated in the extra dimension 
\cite{Bergman:2008sg,Johnson:2008vna}.  

In this appendix we simply give the equations of motion and the free energy for the chirally restored phase
without discussing the solutions. 
We do so for the sake of completeness but also because these expressions may be useful to 
compute possible small corrections to $T_c$ of the order of $N_f/N_c$. One might then speculate whether 
these corrections persist for smaller and thus more realistic values of $N_c$. We leave such a study for the future.

The derivation of the equations of motion and the free energy of the chirally restored phase is analogous to the one for 
the confined phase given in Sec.\ \ref{confinedphase} and Appendix \ref{Appeqs}. The only difference is the use of the metric 
(\ref{dsD8deconf}) instead of (\ref{dsD8conf}) and Eq.\ (\ref{T}) instead of (\ref{MKK}). We use the same coordinate 
transformation as in the chirally broken phase, i.e., Eq.\ (\ref{z}) with $u_{\rm KK}$ replaced by $u_T$ and with $z\in [0,\infty]$. 
This is not really a simplification in this
case but it helps to compare the result to the one for the chirally broken phase. 
We find for the equations of motion 
\be \label{magn1}
\partial_z[k_3(z)\partial_z \hat{b}]=\partial_z[k_3(z)\partial_z b]=0 \, , 
\ee
and 
\begin{subequations} \label{diffeqs1}
\bea
\partial_z[k_0(z)\hat{F}_{z0}] &=& \frac{\alpha M_{\rm KK} u_T^2}{(2\pi T)^3}\left[b(z)F_{z3}+\hat{b}(z)\hat{F}_{z3}\right]\, , \\
\partial_z[k_0(z)F_{z0}] &=& \frac{\alpha M_{\rm KK} u_T^2}{(2\pi T)^3}\left[b(z)\hat{F}_{z3}+\hat{b}(z)F_{z3}\right] \, , \\
\partial_z[k_3(z)\hat{F}_{z3}] &=& \frac{\alpha M_{\rm KK} u_T^2}{(2\pi T)^3}\left[b(z)F_{z0}+ \hat{b}(z)\hat{F}_{z0}\right]\, , \\
\partial_z[k_3(z)F_{z3}] &=& \frac{\alpha M_{\rm KK} u_T^2}{(2\pi T)^3}\left[b(z)\hat{F}_{z0}+\hat{b}(z)F_{z0}\right]\, . 
\eea
\end{subequations}
In contrast to the confined phase, there are now two different functions appearing for the temporal and spatial components,
\be \label{k03}
k_0(z)\equiv \frac{(u_T^3+u_T z^2)^{3/2}}{z\,u_T^{1/2}} \, , \qquad k_3(z)\equiv z\,u_T^{1/2}(u_T^3+u_T z^2)^{1/2} \, .
\ee
The free energy becomes 
\bea \label{Omegadeconf}
\Omega^{\rm deconf} &=& \Omega_g^{\rm deconf}+ \Omega_b^{\rm deconf} + 
\frac{\kappa(2\pi T)^3}{3 M_{\rm KK}u_T^2}\int_{0}^\infty dz\,\left[-k_0(z)(\hat{F}_{z0}^2+F_{z0}^2)+k_3(z)(\hat{F}_{z3}^2+F_{z3}^2)\right]
\non
&&-\;\frac{2\kappa(2\pi T)^3}{3 M_{\rm KK}u_T^2}\left[k_0(z)(\hat{A}_0\hat{F}_{z0}+A_0F_{z0})-k_3(z)(\hat{A}_3\hat{F}_{z3}+A_3F_{z3})
\right]_{z=0}^{z=+\infty}  \, ,
\eea
where
\begin{subequations}
\bea
\Omega_g^{\rm deconf} &\equiv& \frac{32\kappa(2\pi T)^3}{9(2\pi\alpha')^2u_T^2M_{\rm KK}}\int_0^\infty dz\, z\,u_T^{3/2}(u_T^3+u_Tz^2)^{1/6} \, , \\
\qquad \Omega_b^{\rm deconf}&\equiv& \frac{\kappa(2\pi T)}{M_{\rm KK}}(\hat{\cal B}^2 + {\cal B}^2) 
\int_0^\infty dz\,z\,u_T^{1/2}(u_T^3+u_T z^2)^{-5/6} \, .
\eea
\end{subequations}
Here we have assumed the magnetic field to be constant in $z$, $\hat{b}(z)=\hat{\cal B}$, $b(z)={\cal B}$, which solves Eq.\ (\ref{magn1}). 
We see that at the critical temperature where $2\pi T=M_{\rm KK}$ and thus $u_T=u_{\rm KK}$ the free energy assumes a form very similar to
the one in the confined phase. The only differences are then the functions $k_0(z)$ and $k_3(z)$ (vs.\ the single function $k(z)$ in the confined
phase) and the different integrands in $\Omega_g$ and $\Omega_b$.  

\bibliographystyle{JHEP}
\bibliography{refs1}

\end{document}